\input harvmac
\input epsf
\def\Tr{{\rm Tr}}
\def\np{Nucl. Phys. }
\def\pl{Phys. Lett. }
\def\pr{Phys. Rev. }

\def\prl{Phys. Rev. Lett. }
\def\cmp{Comm. Math. Phys. }

\def\sinh{{\rm sinh}}
\def \cosh{{\rm cosh}}

\lref\motl{L.~Motl, hep-th/9701025.}
\lref\bs{T.~Banks, and N.~Seiberg, \np {\bf B 497}(1997) 41,
hep-th/9702187.} 
\lref\dvsq{R.~Dijkgraaf, E.~Verlinde and H.~Verlinde, \np {\bf B 500}
(1997) 43, hep-th/9703030.}
\lref\dvsqr{R.~Dijkgraaf, E.~Verlinde and H.~Verlinde, \np {\bf 62}
(Proc. Suppl.) (1998) 348, hep-th/9709107}
\lref\wynteI{T.~Wynter, \pl {\bf B 415} (1997) 349, hep-th/970929.}
\lref\bbnt{G. Bonelli, L. Bonora, F. Nesti and A. Tomasiello, 
hep-th/9901093.}
\lref\ghv{S.B. Giddings, F. Haquebord and H. Verlinde, \np {\bf B 537}
(1999) 260, hep-th/9804121.}
\lref\bbn{G. Bonelli, L. Bonora, F. Nesti, \pl {\bf B 435} (1998) 303,
hep-th/9805071.}
\lref\bbng{G. Bonelli, L. Bonora, F. Nesti, \np {\bf B 538} (1999) 100,
hep-th/9807232.}
\lref\wynteII{T.~Wynter, \pl {\bf B 439} (1998) 37, hep-th/9806173.}
\lref\gmI{D.J. Gross and P.F. Mende, \pl {\bf B 197} (1987) 129-134.}
\lref\gmII{D.J. Gross and P.F. Mende, \np {\bf B 303} (1988) 407-454.}
\lref\as{A. Sen, Adv. Theor. Math. Phys. {\bf 2} (1998) 51, hep-th/9709220.}
\lref\ns{N. Seiberg, \prl {\bf 79} (1997) 3577, hep-th/9710009.}
\lref\wt{W. Taylor, \pl {\bf B 394} (1997) 283, hep-th/9611042.}
\lref\bsfs{T.~Banks, W.~Fischler, S.H.~Shenker, and L.~Susskind, \pr
{\bf D55}(1996) 5112-5128, hep-th/9610043.}
\lref\sbgsw{S.B. Giddings and S. Wolpert, \cmp {\bf 109} (1987) 177.}
\lref\sbg{S.B. Gidding, Fundamental strings, in Particles, Strings and
Supernovae, proceedings of the 1998 Theoretical Advanced Study
Institute, Brown University, eds. A. Jevicki and C.-I. Tan (World
Scientific, Singapore, 1989).}
\lref\suss{L.~Susskind, hep-th/9705190.}
\lref\BBPT{K. Becker, M. Becker, J. Polchinski, A. Tseytlin,
Phys. Rev. {bf D56} 3174-3178 (1997), hep-th/9706072.}
\lref\CT{I. Chepelev, A.A. Tseytlin, Nucl. Phys. {\bf B515}73-113
(1998), hep-th/9709087}
\lref\gsw{M.B.~Green, J.H.~Schwarz, E.~Witten, Superstring Theory, 
Cambridge University Press (1987).}
\lref\West{P. West, {\it Introduction to Supersymmetry and
Supergravity}, World Scientific (1986).}
\lref\rtI{A. Restuccia and J.G. Taylor, Phys. Rept. {\bf 174} (1989) 283.}
\lref\rtII{A. Restuccia and J.G. Taylor, \pr {\bf D 36} No. 2 (1987) 489.}
\lref\rtIII{A. Restuccia and J.G. Taylor, \pl {\bf B 177} (1986) 39.}
\lref\gs{M.B. Green and J.H. Schwarz, \np {\bf B 243} (1984) 475.}
\lref\Sch{J.H. Schwarz, Phys. Rept. {\bf 89} No. 3 (1982) 223.}
\lref\hp{E. D'Hoker and D.H. Phong, Rev. Mod. Phys. Vol {\bf 60}
(1998) 917.}
\lref\kvh{I.K Kostov and P. Vanhove, \pl {\bf B 444} (1998) 196.}
\lref\mns{G. Moore, N. Nekrasov and S. Shatashvili, hep-th/9803265.}
\lref\witt{E. Witten, \cmp {\bf 141} (1991) 153.}
\lref\fs{F. Sugino, hep-th/9904122.}


\rightline{{\it In memoriam Patrick Heron}\hfill PUPT-1861}
\rightline{hep-th/9905087}
\vskip -10pt
\Title{}
{\vbox{
\centerline{High energy scattering amplitudes}
\centerline{in matrix string theory}
}} 
\vskip-10pt
\centerline{{\bf Thomas Wynter}\footnote{$^\circ$}{{\tt
thwynter@feynman.princeton.edu}}
}
\centerline{{\it Joseph Henry Laboratories, Princeton University,}}
\centerline{{\it Princeton, NJ 08544, USA}}

\vskip .6in
\centerline{{\bf Abstract}}
\vskip .15in
\baselineskip10pt{
\hskip -19.5pt High energy fixed 
angle scattering is studied in matrix string theory. The saddle point
world sheet configurations, which give the dominant contributions to
the string theory amplitude, are taken as classical backgrounds in
matrix string theory. A one loop fluctuation analysis about the
classical background is performed. An exact treatment of the fermionic
and bosonic zero modes is shown to lead to all of the expected
structure of the scattering amplitude. The ten-dimensional Lorentz
invariant kinematical structure is obtained from the fermion zero
modes, and the correct factor of the string coupling constant is
obtained from the abelian gauge field zero modes. Up to a numerical
factor we reproduce, from matrix string theory, the high energy limit
of the tree level, four graviton scattering amplitude. 
}
\bigskip

\Date{May 1999}

\baselineskip=16pt plus 2pt minus 2pt
\bigskip

\newsec{Introduction}

Weakly coupled perturbative string theory is conjectured to be
described by the strong
coupling limit of the matrix string supersymmetric gauge theory
\motl\bs\dvsq \ (see also \dvsqr\ for a review). In 
\dvsq\ it is observed that the strong coupling
limit corresponds to the infra red limit of the theory, and that
symmetry more or less constrains the low energy effective action to be
that of the light cone type IIA CFT. 
Matrix string theory, however, is a non-perturbative description of string
theory and will contain phenomena, inaccessible to the infra red fixed
point reasoning of \dvsq , which can only be investigated by studying
the gauge theory.  A necessary first step in this direction is to
derive, directly from the gauge theory, standard string perturbation
theory. Although this program has not yet been realised several pieces
of the puzzle have been found.

Matrix string theory has classical
solutions corresponding to interacting string world sheets
in light-cone gauge \wynteI (see also \bbnt ). 
The strings split and join via instanton like field configurations
consisting of two regions; a core region, around the interaction
point, where the fields do not commute, and an asymptotic region, away from the
interaction point, where the fields do commute and where the eigenvalues glue
together to form the Riemann surfaces of light-cone string theory
\ghv (see also \bbn ). Arguments have been given that in the large
$N$ \wynteI\ or the strong coupling limit \bbng\ the effective theory
is indeed the type IIA CFT defined on the corresponding Riemann
surface. However there are subtleties that mean that these arguments
are not completely justified. An analysis of the effective expansion
parameter for a perturbative gauge theory loop expansion
\wynteII\ shows that the loop expansion diverges for most physical
scattering processes. There is thus, at present no rigorous approach
to calculating the strongly coupled gauge theory in most situations of
physical interest.

A simple analysis does, however, lead to an explanation of the
factors of the string coupling constant $g_s$ associated with a string
diagram \bbng . In the strong coupling limit there is, in addition to the
Green-Schwarz light cone string action, a completely decoupled abelian
gauge theory. The power of $g_s$ is entirely due to the zero mode
structure of this abelian gauge-field; it is just the difference between
the number of closed, non-exact one-forms (the gauge field zero modes)
and the number of ghost zero modes (exactly one, the constant mode).

In this paper we focus on a scattering process where perturbative
gauge theory calculations can be justified. The scattering process
chosen is high energy, fixed angle, string scattering, originally
studied, from the string theory point of view, in \gmI\gmII\ and
re-analyzed from the perspective of matrix string theory in \ghv
. Indeed it was originally pointed out in \ghv\ that this is a process
where a perturbative gauge theory calculation might be justified. The
scattering amplitude for high energy, fixed angle scattering is
dominated by a classical world sheet configuration
whose size, in the target space, is proportional to the incoming
momenta. In other words high momenta correspond to large (in the
target space sense) world sheets. By going to sufficiently high
momenta, i.e. by making the world sheets sufficiently large, we can
justify using a perturbative gauge theory calculation. This is similar
to matrix theory
\bsfs\ graviton scattering calculations where one pulls the gravitons
sufficiently far apart to be able to use a perturbative calculation.


We will study explicitly the tree level contribution to the four
ground state scattering amplitude.  Starting from the instanton-like
classical matrix string configuration for such a world sheet we will
estimate the fluctuation determinant and calculate exactly the
contribution of the bosonic and fermionic zero modes. We will see that
the zero mode calculation leads to all of the expected structure of
the scattering amplitude. In addition to the correct power of $g_s$
(observed in \bbng ) we will reproduce the precise ten dimensional
Lorentz invariant kinematic structure of the amplitude. The
fluctuation determinant can not be calculated exactly, but we can
nevertheless make some qualitative assessment of its overall form. The
final result being that we will reproduce, up to an unknown numerical
factor from the determinant, the four graviton scattering amplitude.

Although we explicitly focus on the four graviton scattering amplitude
the formalism is general and can be applied to the scattering of any
ground state particles. The calculation can also be generalized, in
principle without difficulty, to loop scattering amplitudes, although
we do not pursue this in this paper. 

We begin, in section 2, by covering some of the essential material
necessary for the calculations of later sections. We start with a very
brief review of matrix string theory. We then turn to saddle point world
sheets in high energy string scattering and how they occur in light
cone string theory. Next we review the instanton
like field configurations corresponding to finite string interactions and
how they embed into a solution corresponding to a string worldsheet
splitting and joining at the interaction points. Finally we study the
kinematics of four particle scattering in light cone gauge in the
centre of mass frame and derive a few simple kinematic identities
useful for section 8. 

In this paper we perform a one loop fluctuation calculation around
classical world sheet configurations of matrix string theory. Section
3 discusses the validity of such a calculation and defines the
Euclidean one loop action. In particular the Minkowski space
Majorana-Weyl fermions are combined into complex fermionic
coordinates and momenta which then allow a Wick rotation into
Euclidean space.

Section 4 analyses the fluctuation determinant. Sections 5 and 6 are
devoted to the zero modes. In section 5 we focus on the behaviour of
the zero modes in the neighbourhood of the instanton-like
configurations, the objective being to understand how the core of the
instanton, where the fields do not commute, effects the finiteness of
the zero mode field configurations. Far from the core of the instanton
these field configurations commute with each other and can be glued
into global zero mode configurations for the global classical field
configurations corresponding to light-cone string diagrams. These
global zero modes we construct in section 6.

In section 7 we discuss the incoming and outgoing wavefunctions and
show how to construct ground state wavefunctions. In particular we
explicitly construct from the fermionic coordinates of section 3 the
graviton wavefunction.

Section 8 is then devoted  to the four graviton scattering  amplitude,
we explicitly integrate over all zero  modes, the integration over the
gauge and ghost zero modes  leading to a  factor of $g_s^2$ (as argued
by \bbng ) and the integration over the fermion zero modes reproducing
the ten dimensional Lorentz invariant kinematic factors  for   four
graviton scattering.

Having focused in previous sections on four
graviton scattering we summarize in section 9 the general procedure
for calculating high energy scattering amplitudes in matrix string
theory, and point out the close connection with standard light cone
gauge superstring calculations. We finish section 9 by discussing how
the analysis might be performed away from the high energy limit, and
speculate on how the calculations could be be given a rigorous
basis. 

Technical details of the Euclideanization of the fermions and
the fermion zero mode integrations are contained in the appendix.

\newsec{Matrix string theory, world sheets and high energy scattering}
\subsec{Matrix String Theory}
Matrix string theory \motl\bs\dvsq\ is equivalent to ten
dimensional SYM theory dimensionally reduced to two dimensions~:
\eqn\mst{
S={1\over 2\pi}\int\,d\tau d\sigma\Tr\bigl[
-{1\over 2}(D_{\alpha}X^I)^2+{i\over 2}S^T\slash \hskip -6.5pt DS
-{1\over 4}F_{\alpha\beta}^2+{1\over 4g_s^2}[X^I,X^J]^2
+{1\over 2g_s}S^T\Gamma_I[X^I,S]\bigr].
}
A careful derivation of this action (using the ideas of \as\ns\wt )
from the original matrix theory proposal \bsfs\ can be found in \ghv .
All fields are $N\times N$ hermitean matrices. The index $I$ for the
bosonic fields runs from 1 to 8 and corresponds to the transverse
directions of the ten dimensional target space. The 16 component
fermion fields $S$ split into $S^a$ and $S^{\dot{a}}$, the $8_s$ and
$8_c$ representations of $SO(8)$. The matrices $\Gamma^{\mu}$,
$\mu=1,\cdots,9$ are the $spin(8)$ gamma matrices with $\Gamma^0$ in
the Dirac operator $\slash \hskip -6.5pt D$ equal to the sixteen by
sixteen unit matrix. $g_s$ is the string coupling constant and the
coordinate $\sigma$ runs from $0$ to $2\pi$. The action \mst\ is
conjectured to describe non-perturbative type IIA string theory
compactified on a light-like circle with $N$ the number of quanta of
$p^+$ momenta along the compactified direction.

String world sheets are described in matrix string theory by commuting
matrix configurations in which the $N$ eigenvalues of the matrices
form a branched covering of the cylinder, and hence form the Riemann
surfaces of interacting light-cone string theory. These surfaces are
characterized by strings of different length which split and join, the
total length of the strings, which corresponds to the light-cone $p^+$
momentum, being preserved. In the limit $N\rightarrow\infty$ the
moduli space of the branched coverings is equivalent to the moduli
space of all possible two-dimensional Riemann surfaces. A crucial
ingredient of these configurations is that they are associated with a
topologically non-trivial two dimensional gauge field, which via
Wilson lines generates the correct monodromy around all the branch
points \wynteI . Specifically a single valued matrix description of a
multivalued branched covering of the cylinder is given by
\eqn\brcv{
X=U{\rm diag}(x_1,\cdots ,x_N)U^{\dagger},\quad 
A_{\alpha}=ig_s U^{\dagger}(\partial_{\alpha}U),
}
where the eigenvalues $x_i$ form the branched covering and the unitary
matrix $U$ (which is multivalued) generates the monodromies around
the branch points. 

The gauge field configurations however are singular at the branch
points and lead to a delta function singularity in the field
strength. As was realised in \ghv\ this singularity can be resolved by
instanton like field configurations. Far from the core of the
instanton the fields commute with each other and their eigenvalues
describe a Riemann surface with a branch point. Close to the core of
the instanton, however, the fields no longer commute and the Riemann
surface interpretation breaks down. It is to be expected that any
classical string world sheet will have a corresponding classical
solution in matrix string theory. 
\vskip 20pt
\hskip 26pt\epsfbox{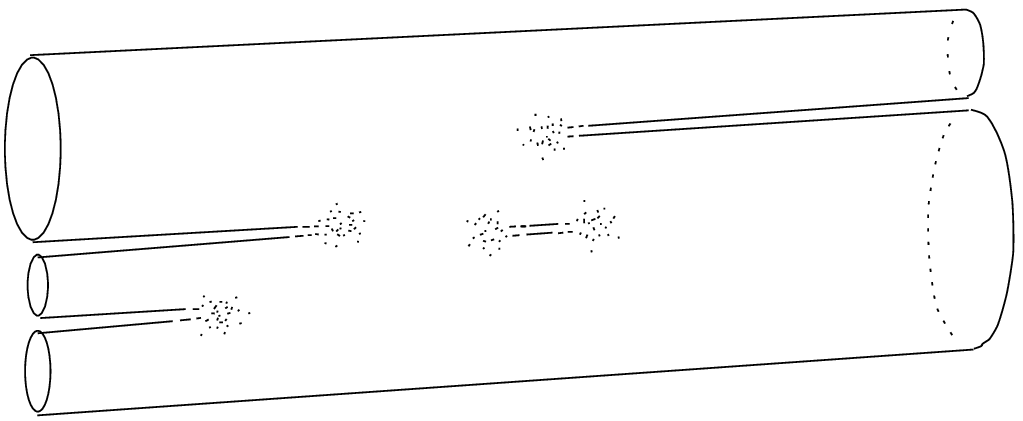}
\vskip 20pt
\centerline{{\bf Fig 1}. 
Light cone string world sheets in matrix string theory}
\vskip 20pt

In \ghv\ it was shown that there exists a physically very interesting
class of classical world sheets that preserve a certain amount of
symmetry around the branch points and hence allow a simple
construction of the instanton like field configurations. These world
sheets are the saddle point classical world sheets
dominating the amplitudes of high energy, fixed angle string
scattering.  Before reviewing the construction of the matrix
string solutions we briefly recall how these saddle point world sheets
appear in light cone string theory.

\subsec{High energy scattering and light cone string theory}
High energy, fixed angle scattering processes were studied in string
theory in \gmI\gmII . String theory simplifies enormously in this
limit as the integral over world sheets localizes around a finite
number of saddle-point configurations. It was conjectured in
\gmI\gmII\ that the
saddle point world sheets of different genus follow the same target
space path, up to an overall scaling factor.  This simplification
permits one to study string perturbation theory out to arbitrarily
high orders. One of the results of this analysis is that the
perturbation series is extremely divergent, with the genus $g$ world
sheets contributing a factor of $g^{9g}$. A non-perturbative
description such as matrix string theory should provide a natural
cut-off to this divergence. First steps in an analysis along these
lines were taken in \ghv . We will be less ambitious in this article
and focus on retrieving from matrix string theory standard string
perturbation theory. For the purposes of this article, we focus on
high energy scattering as a means to justify using a perturbative
gauge theory calculation. Below we describe how the saddle-point world
sheets arise in light cone string theory.

In light cone string theory the Virasoro
constraints have been imposed at the classical level to eliminate the
non-physical degrees of freedom. Specifically light cone string 
coordinates, $X^{\pm}=1/\sqrt{2}(X^0\pm X^9)$ have been eliminated
from the dynamics by imposing the gauge
\eqn\Xplus{
X^+=\tau,
}
where $\tau$ is the world sheet time direction. The coordinate $X^-$
is then a function of the transverse coordinates $X^I$
($I=1,\cdots,8$) and the fermion fields.  

The light cone coordinates $w=\tau+i\sigma$ are defined using the
Mandelstam mapping from the light cone diagram to the Riemann surface
with uniformization $z$
\eqn\unif{
w=\sum_ip_i^+G(z,z_i),}
where $G$ is an abelian differential with purely imaginary periods and
real part equal to the Greens function. The existence and uniqueness
of such a differential for arbitrary genus has been proved in
\sbg\sbgsw .  The branch points of the light cone string diagram are
situated at the zeros of $\omega$, i.e. the stationary points of $w$
as a function of $z$. In the neighbourhood of a simple zero of
$\omega$ situated at $z_0$ we have $w-w_0=(z-z_0)^2$.

For tree level scattering we have
\eqn\sphwz{
w=\tau+i\sigma=\sum_i\epsilon_iN_i\log(z-z_i),
}
with $z$ defined in the complex plane and
$\epsilon_iN_i$ the $p^+$ momentum of the $i$th string. The
$\epsilon_i$ are equal to $+1$ for incoming states and $-1$ for
outgoing states. The lengths of the strings are proportional to their
$p^+$ momenta. The light-cone interaction points are given by the
zeros of $\partial_zw=0$, i.e. by the roots of the polynomial equation
\eqn\rts{
\sum_i{\epsilon_i N_i\over z-z_i}=0.
}
The classical solution for a string world sheet is given by
\eqn\wdst{
X=\sum_i\epsilon_ip_i\log|z-z_i|,
}
where $p_i$ are the transverse momenta and $z=z(w)$ is defined through
\sphwz .

For the classical field configurations \wdst\ the light cone 
action is given entirely by boundary terms
\eqn\clasS{
S=-{1\over 4\pi}\int d^2w (\partial_{\alpha}X^I)^2
={1\over 4\pi}\sum_i\epsilon_i\oint d\sigma
X^I\partial_{\tau} X^I\bigl|_{\tau=\tau_i},
}
where the $\tau_i$ are the initial(final) times for the scattering
process and the $\sigma$ integral for the $i$th string runs over an
interval of length $N_i$. For $\tau_i$ very big, i.e. far from the
interaction region, equation \sphwz\ can be inverted perturbatively
and substituted into \wdst\clasS\ to obtain the results
\eqn\classS{\eqalign{
X=&{p_i\over N_i}\tau
+\sum_{i\neq j}\epsilon_j\bigl[p_j-p_i{N_j\over N_i}\bigr]\log|z_i-z_j|
+\cdots\cr
S=&\sum_i \epsilon_ip^-_i\tau_i\,\,+\,\,
{1\over 2}\sum_{i\neq j}\epsilon_i\epsilon_j
\bigl[p^I_ip^I_j-p^-_iN_j-N_ip^-_j\bigl]\log|z_i-z_j|
\,\,+\,\,\cdots,}}
where the $p^-$ momentum is given in terms of the transverse momenta
$p_i$ and $p^+$ momentum $N_i$ by
\eqn\defpm{
p^-_i={|p^I_i|^2\over 2N_i}
}
The dots in \classS\ correspond to exponentially small
corrections. The first term in the action \classS\ is a phase factor
corresponding to the time evolution of the incoming and outgoing states
with respect to light cone Hamiltonian $p^-$, and cancels out in
scattering amplitudes (see
\gsw ). The second term is the physically relevant Lorentz invariant
contribution~:
\eqn\classSmink{
{1\over 2}\sum_{i\neq j}k_i.k_j\log|z_i-z_j|,
}
where the $k_i$ are ten momenta defined to be incoming to the
scattering process, i.e. $k_i=(N_i,\,|p^I_i|^2/(2N_i),\, p^I_i)$ for
incoming states and $k_i=-(N_i,\,|p^I_i|^2/(2N_i),\, p^I_i)$ for outgoing
states. 

The saddle point contribution to the action is found by looking for
the stationary point of the above term under variations with respect to
the modular parameters $z_i$.

\subsec{Classical world sheets in matrix string theory}

The matrix string solutions for a general classical world sheet can be
expected to be very complicated and to date no exact solution for a
matrix string world sheet with asymptotic states has been constructed
\footnote{$^1$}{Exact matrix string world sheets have been
constructed and studied in depth in \bbnt\bbng , however these are for
classical world sheets which are entirely holomorphic or entirely
anti-holomorphic functions of the world sheet coordinate and thus
cannot represent the matrix string versions of the classical world
sheets \wdst . We thus do not consider them further in this
article.}. A successful strategy was however developed for the saddle
point world sheet configurations which dominate high energy scattering
amplitudes (see \ghv ). Starting with the saddle point world sheet the
authors of \ghv\ first identified the type of square root (holomorphic
and/or anti-holomorphic) involved at the interaction points. They then
constructed instanton-like field configurations whose behaviour far
from the core matched the square root behaviour of the world sheet. 
The approach is an approximation, in that the instanton-like field
configurations have a finite size and should strictly be matched onto
more than just the dominant square root dependence on the world
sheet coordinates. The true instanton-like field configuration would
thus be expected to be slightly modified away the core. 

In \ghv\ this program was implemented explicitly for the construction
of the four string scattering process. Below we briefly review the
main points. In the centre of mass frame the scattering process
describes a two dimensional plane and the corresponding classical
world sheet can be described by just two of the eight transverse
coordinates\footnote{$^2$}{Scattering processes involving more
than four particles would require three or more transverse coordinates
to describe their classical world sheets.}, $X^1$ and $X^2$. For
convenience these are combined into a pair of complex fields defined
by
\eqn\defX{
X={1\over\sqrt{2}}(X^1+iX^2)\quad{\rm and}\quad 
X^*={1\over\sqrt{2}}(X^1-iX^2).
}
The light cone world sheet has two branch points $z_0^+$ and $z_0^-$
given by the two roots of the quadratic equation for four particle
scattering of \rts .
The dominant classical world sheet has the property that the 
complex scalar field $X$ is an anti-holomorphic, respectively, holomorphic
function of $z$ in the neighbourhood of the two branch
points. Specifically~:
\eqn\holbpt{
\partial_{z}X|_{z=z_0^+}=
\sum_i{\epsilon_i p_i\over z-z_i}|_{z=z_0^+}=0,
}
and
\eqn\antiholbpt{
\partial_{\bar{z}}X|_{z=z_0^-}=
\sum_i{\epsilon_i p_i\over \bar{z}-\bar{z}_i}|_{z=z_0^-}=0.
}
This symmetry allows a simple construction for the instanton-like
field configurations around the branch points. The starting point is
the four dimensional self dual YM equation dimensionally reduced to 
two dimensions. The two possible signs for the self dual equations
correspond to the two different holomorphic behaviours of equation
\holbpt .
The instanton solution around the branch point $z_0^+$
is obtained from 
$F_{\mu\nu}=\epsilon_{\mu\nu\rho\sigma}F_{\rho\sigma}$ with
$\epsilon_{0129}=+1$. In terms of the two dimensional fields the
equation reads
\eqn\sfdl{\eqalign{
F_{w\bar{w}}=&-{i\over g_s}[X,X^*]\cr
D_wX=&0\cr
D_{\bar{w}}X^*=&0}}
where, in addition to the complex scalar fields defined in \defX\ 
we have introduced complex two dimensional gauge fields and coordinates
\eqn\defA{
A={1\over\sqrt{2}}(A^0-iA^9)
\quad{\rm and}\quad
w={1\over\sqrt{2}}(\sigma^0+i\sigma^9).
}

A simple solution to these equations of motion corresponding to a
simple branch point involving two eigenvalues is 
\eqn\hatX{
X=U\hat{X}U^{\dagger}
\quad{\rm and}\quad 
A=ig_sU^{\dagger}\partial U
\quad{\rm with}\quad
\hat{X}=B\sqrt{\bar{w}}\tau_3,
\quad{\rm and}\quad 
U=e^{{1\over 8}\ln{w\over\bar{w}}\tau_1} .
}
As already described
this leads to a delta function singularity in the field strength at the
interaction points. This singularity
can be removed once we are working with complex
coordinates $X$, by using a complexified ``gauge'' transformation $G$
which also has a singularity at the origin tuned in such a way as to
leave a singularity free field strength. 
With this insight the solution can be written in the form
\eqn\instsol{\eqalign{ X=&UG\hat{X}
G^{-1}U^{\dagger}\cr
A=&-ig_s\bigl[G^{-1}(\partial_wG)+U^{\dagger}(\partial_wU)\bigr],}} 
where the diagonal matrix $\hat{X}$ and unitary matrix $U$ are 
as in \hatX\ and the matrix $G$ is
given by 
\eqn\Gdef{ 
G=e^{\alpha(w\bar{w})\tau_1}.}  
This ansatz
automatically satisfies the last two equations of \sfdl\ with the
first equation leading to a differential equation for $\alpha$ 
\eqn\difa{ 
(\partial_r^2+{1\over r}\partial_r)\alpha= 
{8B^2\over g_s^2}r\,\sinh {2\alpha} 
\quad{\rm with}\quad
\alpha\rightarrow\cases{0\quad{\rm for}\quad r\rightarrow\infty\cr
-{1\over 4}\ln{r}\quad{\rm for}\quad r\rightarrow 0} } 
where
$r=\sqrt{w\bar{w}}$ is the radial distance from the branch point.  The
boundary conditions are necessary for a finite solution. In
particular the second boundary condition ensures that there are no
${1\over w}$ pole terms in the gauge field $A$ and hence no delta
function singularity in the field strength $F_{w\bar{w}}$.

There is no explicit solution to this equation but it can easily be
solved numerically. In terms of the function $\alpha$ the expressions
for the scalar fields $X$ and the field strength $F_{w\bar{w}}$ are
given by the simple expressions 
\eqn\XF{
X=B\sqrt{\bar{w}}(\cosh\alpha\,\tau_3\,+\,i\sinh\alpha\,\tau_2)
\quad{\rm and}\quad
F_{w\bar{w}}={2iB^2\over g_s}r\,\sinh{2\alpha}.
}
For simplicity we do not include the final gauge transformation $U$

The instanton like solutions \XF\ are embedded into a global solution,
defined on the cylinder, corresponding to the  classical
world sheet. The parameter $B$ of equations \hatX ,\difa\ and \XF\
is determined in terms the four external momenta by
\eqn\defB{
|B|^2={|p_1p_3^*-p_1^*p_3|(N_1+N_2)
\over\sqrt{N_1N_2N_3N_4}},
}
where the $p_i$ and $p_i^*$ are the complex transverse momenta
associated with the field $X$ and its complex conjugate. The $N_i$ are
the $p^+$ momenta for the $i$th string.

Finally we note that there are two physical scales associated with the 
instanton. Firstly the differential equation \difa\ can be given a 
dimensionless form by
absorbing the coupling constants into a rescaling of the radial
coordinate. In other words the instanton has a natural world sheet
scale~: 
\eqn\lins{
l_{inst}\sim\biggl({g_s^2\over B^2}\biggr)^{1\over 3}.
}
Secondly, a simple analysis of the limiting behaviour of the field $X$
at the origin shows that there is a minimal target space ``distance'' 
between the two strings~:
\eqn\dins{
d_{inst}=\sqrt{\Tr(X(0)X^*(0))}\sim g_s^{1\over 3}|B|^{2\over 3}.
}

\subsec{Kinematic relations for four string scattering}
We will explicitly study the four string ground state scattering
process. The calculation is performed in the centre of mass frame. In
this subsection we give various kinematic identities useful for the
calculations of later sections. 

All particles are massless. They have transverse momentum lying in the
$X^1$,$X^2$ plane specified by the complex numbers
$p_i=1/\sqrt{2}(p_i^1+ip_i^2)$. The $p^+$ momentum is given by the length
$N_i$ of the string. To make contact with standard conventions for
scattering processes we define all ten-vectors of momenta $k_i$ to be
incoming momenta~:
\eqn\mom{
k_i=(p^+_i,p^-_i,p_i,p_i^*)
=\cases{(N_i,{|p_i|^2\over N_i},p_i,p_i^*)\quad{\rm for}\quad i=1,2\cr
(-N_i,-{|p_i|^2\over N_i},-p_i,-p_i^*)\quad{\rm for}\quad i=3,4\cr},}
where
\eqn\bpb{
|p_i|^2=p_i p_i^*={1\over 2}((p_i^1)^2+(p_i^2)^2).}
The dot product of two momenta then reads
\eqn\pdotp{
k_i.k_j=-p^+_ip^-_j-p^+_jp^-_i+(p_ip_j^*+p_i^*p_j).}
In the centre of mass frame conservation of momentum 
reads
\eqn\consm{
\eqalign{
&p_1+p_2=p_3+p_4=0\cr
&N_1+N_2=N_3+N_4=N\cr
&|p_1|^2 {N\over N_1N_2}=|p_3|^2 {N\over N_3N_4}=p^2,\cr
}
}
where in the final line (conservation of $p^-$ momentum) we have used 
the first and second lines (conservation of transverse and $p^+$ 
momenta) to simplify the 
result and to define the quantity $p^2$. Inverting the third line we 
can write $|p_1|^2$ and $|p_3|^2$ in terms of $p^2$
\eqn\poptp{ 
|p_1|^2={N_1N_2\over N}p^2
\quad\quad{\rm and}\quad\quad
|p^3|^2={N_3N_4\over N}p^2.}
Finally it is useful to define the ten dimensional Lorentz invariant 
quantities for the scattering process
\eqn\stu{\eqalign{
&s=-(k_1+k_2)^2=-2k_1.k_2\cr
&t=-(k_2+k_3)^2=-2k_2.k_3\cr
&u=-(k_1+k_3)^2=-2k_1.k_3,\cr}
}
where the sign conventions are, as before, for $k_1$,$k_2$,$k_3$ and $k_4$
incoming momenta. Using the definitions for the  
$k_i$ \mom\ along with the identities \consm\ and \poptp\ the Lorentz
invariants $s$,$t$ and $u$ can be written as
\eqn\stuexp{
\eqalign{
&s=\,\,2p^2\cr
&t=-2{(N_1N_3+N_2N_4)\over N^2}p^2-2q^2\cr
&u=-2{(N_1N_4+N_2N_3)\over N^2}p^2+2q^2\cr
}
\quad\quad{\rm with}\quad\quad
q^2=p_1^*p_3+p_3p_1^*.
}
$p^2$ in the above expressions is given in \consm\poptp . It is easy
to check that the above expressions satisfy the identity relating $s$,
$t$ and $u$ for massless particles
\eqn\stuid{
s+t+u=0.}
\newsec{Perturbative calculations and the Euclidean action}

There are two standard limits one can take in matrix
string theory to recover perturbative string theory.  Firstly one can
take the limit $N\rightarrow\infty$ followed by $g_s\rightarrow 0$. By
the matrix string theory conjecture this is the limit necessary
to recover perturbative type IIA string theory in uncompactified ten
dimensional space. Secondly one can hold $N$ finite and send
$g_s\rightarrow 0$. By Susskind's conjecture \suss\ this limit
corresponds to string theory compactified on a light-like circle with
$N$ the number of $p^+$ quanta around the compact direction. 

Since both are strong coupling limits one would not expect
perturbative gauge theory calculations to be justified. Indeed, a
perturbative gauge theory expansion would involve, through the three
and four point vertices, arbitrary powers of $1/g_s$ which diverge as
$g_s\rightarrow 0$. However, when deciding whether or not a
perturbative calculation is justified it is the effective parameter
weighting the loop expansion that is important not the size of the
coupling constants. This is determined by both the coupling constants
and the masses of the particles exchanged in the propagators. The
masses of the quantum fluctuations are determined by the commutator
term in the action \mst\ and are hence proportional to $1/g_s$. The
mass of the propagators can thus compensate for the strength of the
coupling constants. A good analogy to bare in mind is the top quark and
Higgs in the standard model. The coupling of the top quark to the
Higgs field is enormous (it is this that gives it its enormous
mass). This does not mean that perturbative standard model
calculations involving a virtual top quark and Higgs are not justified
(ask any phenomenologist).  There is precise cancellation between the
mass in the propagator of the virtual top quark and the coupling of
that virtual top quark to the Higgs. An identical cancellation happens
in matrix(string) theory.

In the context of matrix string theory an analysis of the balance
between these two effects, in the large $N$ limit, was carried out in
\wynteII . Starting from a bosonic background field there is a 
systematic expansion in powers of derivatives of the background fields
for the effective action that one can calculate. It's overall form is
entirely determined by dimensional reasoning and the identification of
the loop counting parameter \BBPT\CT . The result is that the
effective action can be written in the form
\eqn\Leff{
S_{\rm eff} = \int d^2\sigma 
\biggl[F^2 + \sum_{L=1}^{\infty}{\cal L}_L \biggr]
\quad{\rm with}\quad
{\cal L}_L =\sum_{n=2}^{\infty} g_s^{2n-2} {F^{2n}\over X^{4n+2L-4}}.
}
where $F^{2n}/X^{2m}$ means bosonic terms with $2n$ derivatives in the
numerator and $2m$ powers of the scalar fields $X^I$ in the denominator.

The analysis of \wynteII\ showed that the two standard limits
mentioned above lead to divergent loop expansions. For the case
$N\rightarrow\infty$ before $g_s\rightarrow 0$ the tensor structures of
the terms $F^{2n}/X^{4n+2L-4}$ are such that the loop expansion
diverges with positive powers of $N$. The physical cause of the
divergence is that neighbouring strips (eigenvalues) of the background
long string configuration come arbitrarily close together in the limit
$N\rightarrow\infty$. The off diagonal field variables connecting the
neighbouring strips thus become massless in this limit.  Holding $N$
finite and sending $g_s\rightarrow 0$ leads to a different kind of
problem. Naively from \Leff\ the effective action would be well
behaved in this limit.  However close to the interaction points the
instanton-like field configurations depend upon $g_s$. Taking this
into account leads to a loop expansion weighted by inverse powers of
$g_s$, (see \wynteII ) which diverges in the limit 
$g_s\rightarrow 0$. Physically the reason for this divergence is that,
as can be seen from \dins , the ``minimal distance'' between the two
strings tends to zero in the limit $g_s\rightarrow 0$. This again
leads to massless fluctuations and a divergence.  To be able to
calculate in these two limits would thus first require the development
of direct strong coupling techniques.

In this article we will avoid these problems by studying a third limit
which will allow us to calculate a perturbative string theory
scattering amplitude by performing a perturbative gauge
theory calculation. The limit corresponds to high energy string
scattering in the finite $N$ version of matrix string
theory. Specifically we will hold $N$ fixed, hold $g_s$ fixed and
study the dominant contribution in the limit in which the external
momenta, $p_i$, of the scattering process tend to infinity. Note that
$N$ can be large and $g_s$ can be small. What is important is that we
send $p_i\rightarrow\infty$ before taking any other limit.

The strategy of this paper is to perform a fluctuation calculation
about a classical matrix string theory background corresponding to one
of the dominant saddle point world sheets found by Gross and Mende
\gmI\gmII\ in the
$p_i\rightarrow\infty$ limit. As can be seen directly from \wdst\ the
target space size of the saddle point backgrounds are proportional to
the external momenta. $p_i\rightarrow\infty$ is thus a limit in which
the target space size of the background becomes infinite. Looking at
loop expansion \Leff\ we see that there are always more powers of the
background field in the numerator than in the denominator. In the
limit in which the background becomes infinitely large these terms
will thus be scaled away altogether. Physically the limit has the
effect of separating, in target space, the individual
strips(eigenvalues) of the matrix background, with the result that the
off-diagonal elements connecting together the strips become infinitely
massive. Again this simple argument could be invalidated by the
existence of the interaction points where two eigenvalues are
connected by a branch point and come close together. As already stated
the instanton-like field configurations lead to their being a
``minimal distance'' \dins\ between the two eigenvalues. Reading off
directly from \dins\defB\ we see that this minimal distance tends to
infinity if we take the limit $p_i\rightarrow\infty$ while holding $N$
and $g_s$ fixed. There is thus no problem of massless fluctuations as
there was for the $g_s\rightarrow 0$ limit.

To summarize, taking the limit $p_i\rightarrow\infty$, before taking
any other limit will lead to a loop expansion in which the
contributions from higher order loops will be scaled away. It is thus
justified in this limit to use a one loop calculation. As will be
discussed below the calculation reduces to the evaluation of the
determinant for the quadratic fluctuations (by the above arguments
this will be effectively equal to one) and to an
integration over the zero modes.

A final comment is in order. Asymptotically
far from the interaction points the classical background splits into
separate blocks for the different strings. Each block is proportional
to a unit matrix. The off diagonal elements within a block are thus
massless. This is a result of the fact that we are not using true
matrix string wavefunctions. Presumably there is some LSZ type
reasoning that could be developed to justify replacing the true matrix
string wavefunctions by simple diagonal blocks. 

\subsec{Euclidean fermions}
The fermions of the action \mst\ are Majorana Weyl fermions in ten
dimensional Minkowski space and consist of sixteen real fermionic
components. We will need to work in ten dimensional Euclidean space
(dimensionally reduced to two dimensions)
where the instanton configuration and functional integral are defined,
and where it is not possible to define Majorana Weyl spinors. 
In addition the Majorana Weyl fermions satisfy the commutation
relations
\eqn\mwfcomm{
\{S^a(\sigma),S^b(\sigma')\}
=\pi\delta^{ab}\delta(\sigma-\sigma')
\quad{\rm and}\quad
\{\tilde{S}^{\dot{a}}(\sigma),\tilde{S}^{\dot{b}}(\sigma')\}
=\pi\delta^{\dot{a}\dot{b}}\delta(\sigma-\sigma'),
}
where $a,b$ and $\dot{a},\dot{b}$ denote respectively the ${\bf 8_s}$
and ${\bf 8_c}$ representations of $SO(8)$. In other words the spinor
$S^a$ is simultaneously a coordinate and its conjugate momenta, and
similarly for $\tilde{S}^{\dot{a}}$. The spinor variables can however
be combined into complex spinor coordinates, which are distinct from
their conjugate momenta, and which can be used to define a Euclidean
action. The complexification procedure means that the $SO(8)$ symmetry 
of the theory will no longer be manifest, although, as we will see, it
is restored in the calculation of physical amplitudes. A
similar complexification procedure has to be carried out in 
light cone superstring theory \gsw .  

The calculation of this paper is up to quadratic order only in the
background fluctuations. To quadratic order the Dirac operator
involves only four gamma matrices, $\Gamma^0$ and $\Gamma^9$, which
couple to the background gauge field and $\Gamma^1$ and $\Gamma^2$
coupling to the background fields for $X^1$ and $X^2$.  This fact
permits a simple definition for the Euclidean action.  We start by
choosing a basis for the gamma matrices such that the four gamma
matrices, $\Gamma^0$,$\Gamma^9$,$\Gamma^1$ and $\Gamma^2$, can be
written as four by four blocks (see appendix). We then define four
spinor coordinates
$\theta^A$,$\bar{\theta}_{\bar{A}}$,$\tilde{\theta}^A$,~
$\bar{\tilde{\theta}}_{\bar{A}}$
and four corresponding momenta 
$\lambda_A$,$\bar{\lambda}^{\bar{B}}$,$\tilde{\lambda}_A$,~
$\bar{\tilde{\lambda}}^{\bar{A}}$ 
by (see appendix)
\eqn\defth{
(v\otimes v\otimes1)S^a=
\pmatrix{\theta^A\cr
\bar{\lambda}^{\bar{A}}\cr
\bar{\theta}_{\bar{A}}\cr
\lambda_A\cr}
\quad{\rm and}\quad
(v\otimes v\otimes1)S^{\dot{a}}=
\pmatrix{\tilde{\theta}^A\cr
\bar{\tilde{\lambda}}^{\bar{A}}\cr
\bar{\tilde{\theta}}_{\bar{A}}\cr
\tilde{\lambda}_A\cr},
\quad{\rm with}\quad
v={1\over\sqrt{2}}\pmatrix{1&i\cr 1&-i\cr}.
}
The indices $A$ and $\bar{A}$ take the values $A,\bar{A}=1,2$.  Note
that $\bar{\theta}$ is not the complex conjugate of $\theta$, they are
distinct variables. The notation is chosen to match standard
conventions for spinors in four dimensional space (see \West ). This
choice of coordinates breaks the $spin(8)$ symmetry of the spinors
down to $U(1)\otimes U(1)\otimes SU(2)\otimes SU(2)$.  The two $U(1)$s
correspond to $SO(2)$ rotations in the $X^1$,$X^2$, and (with the
basis of gamma matrices chosen in the appendix) $X^3$,$X^4$
planes. The $SU(2)\otimes SU(2)$ corresponds to an $SO(4)$ for the
remaining four dimensional space $X^5,X^6,X^7$ and $X^8$. The charges
of the fermions under the rotations in the $X^1$,$X^2$ and $X^3$,$X^4$
planes as well as the transformations under rotations in the
$X^5,X^6,X^7,X^8$ can be read off directly from the form of the eight
dimensional $\Gamma^{IJ}$ (see appendix). The final result of this
decomposition is that we have the following bosonic and fermionic
coordinates.
\eqn\bfcoord{
\eqalign{
&X,X^*={1\over\sqrt{2}}(X^1\pm iX^2)\cr
&\tilde{X},\tilde{X}^*={1\over\sqrt{2}}(X^3\pm iX^4)\cr
&X^m\quad{\rm for}\quad m=5,\cdots,8\eqalign{& \cr & \cr}\cr
}
\quad\quad\quad\quad\quad\quad
\eqalign{
&\theta^A({1\over 2},{1\over 2})\cr
&\bar{\theta}_{\bar{A}}(-{1\over 2},{1\over 2})\cr
&\tilde{\theta}^A(-{1\over 2},{1\over 2})\cr 
&\bar{\tilde{\theta}}_{\bar{A}}({1\over 2},{1\over 2}).\cr
}
}
The first and second arguments of the fermions are their $U(1)$
charges under rotations in respectively the $X^1,X^2$ and $X^3,X^4$
planes. The absence or presence of the bar above the fermions
indicates how they transform under rotations in the $X^5,X^6,X^7,X^8$
space. Specifically under such an infinitesimal rotation we have 
\eqn\bfrot{\eqalign{
\delta X^m = &({1\over 2}w^{mn}J^{mn})^{pq}X^q\cr
\delta \theta^A = &{1\over 4}w^{mn}(\sigma^{mn})^A_{\,\,\,\,B}\theta^B\cr
\delta \bar{\theta}_{\bar{A}} = &
{1\over 4}w^{mn}(\bar{\sigma}^{mn})_{\bar{A}}^{\,\,\,\,\bar{B}}
\bar{\theta}_{\bar B}\cr
}
\quad\quad{\rm with}\quad\quad
\eqalign{
&(J^{mn})^{pq}=\delta^{mp}\delta^{nq}-\delta^{mq}\delta^{np}\cr
&(\sigma^{mn})^A_{\,\,\,\,B}=
{1\over 2}(\sigma^m\bar{\sigma}^n-\sigma^n\bar{\sigma}^m)\cr
&(\bar{\sigma}^{mn})_{\bar A}^{\,\,\,\,\bar{B}}=
{1\over 2}(\bar{\sigma}^m\sigma^n-\bar{\sigma}^n\sigma^m).\cr
}
}
The indices $mnpq$ run from 5 to 8. The $\sigma$ and $\bar{\sigma}$
matrices are given in terms of the pauli matrices by 
\eqn\defss{
(\sigma^m)^{A\bar{B}}=(i,\tau_1,\tau_2,\tau_3)
\quad{\rm and}\quad
(\bar{\sigma}^m)_{\bar{A}B}=(-i,\tau_1,\tau_2,\tau_3).}
Note that this decomposition of the 
fermions into coordinates and momenta differs from
the standard decomposition used in light cone superstring theory
\gs\gsw\ where one breaks the $spin(8)$ symmetry of the spinors into
$U(1)\otimes SU(4)$. The classical background
further breaks the $SU(4)$ down to $U(1)\times SU(2)\times SU(2)$. 

In section 8 we will use the transformation properties of the
spinors $\theta$,$\bar{\theta}$,$\tilde{\theta}$ and 
$\bar{\tilde{\theta}}$ to construct incoming and outgoing wavefunctions
transforming under $SO(8)$ rotations. Specifically we will construct 
the combinations of $\theta$ and $\bar{\theta}$ corresponding to the 
$X$,$X^*$,$\tilde{X}$,$\tilde{X}^*$ and $X^m$ components 
of an $SO(8)$ vector and similarly for $\tilde{\theta}$ and 
$\bar{\tilde{\theta}}$. These ``left'' and ``right''
$SO(8)$ vectors will form incoming and outgoing graviton states. We
will then explicitly calculate the four graviton scattering amplitude.

\subsec{The Euclidean action}
It is convenient in this section to use ten dimensional notation for
the bosonic fields. Indices $\mu$ run from $\mu=0,\cdots,9$ and split
into indices $0,9$ for the two dimensional cylindrical coordinates on
which the fields depend and $I=1,\cdots,8$. All bosonic fields are
denoted by $A^{\mu}$ with $A^I=X^I$. We then split the fields into a
background part, $A$, corresponding to the classical matrix
string solution and a fluctuation part $V$~:
\eqn\AAclV{
A_{total}=A+V,
}
with $V^I=Y^I$.
To calculate the effect of quantum fluctuations about the classical
configuration the action needs to be gauge fixed. We use the
standard background covariant gauge fixing term
\eqn\gfix{ 
{\cal L}_{gf}=(D_{\mu}V_{\mu})^2
\quad{\rm with}\quad
D_{\mu}=\partial_{\mu}+{i\over g_s}[A_{\mu},\,\,].
}
The Euclidean Lagrangian for the quadratic fluctuations reads
\eqn\onelL{
{\cal L}_{Eu}=\Tr\bigl[ -(D_{\mu}V^{\nu})^2
-2{i\over g_s}V^{\mu}[F^{\mu\nu},V^{\nu}]
+\lambda^T\slash \hskip -6.5pt D\theta
+c^*D^2c\bigr],
}
where $V$ are the bosonic fluctuations and $\theta$ and $\lambda$ the 
fermionic coordinates and momenta defined in \defth . The fields 
$c$ are the ghosts. $D$ is the background covariant
derivative given in \gfix . Using the definition of gamma matrices
given in the appendix the fermionic part of the Lagrangian reads
\eqn\fermL{
\lambda^T\slash \hskip -6.5pt D\theta
=\pmatrix{\lambda_A \tilde{\lambda}_B \bar{\lambda}^{\bar{A}}
\bar{\tilde{\lambda}}^{\bar{B}}\cr}
\pmatrix{D_w&{i\over g_s}[X^*,\,\,]& &\cr
-{i\over g_s}[X,\,\,]&D_{\bar{w}}& &\cr
& & &D_w&-{i\over g_s}[X,\,\,]\cr
& & &{i\over g_s}[X^*,\,\,]&D_{\bar{w}}\cr}
\pmatrix{\theta^A\cr\tilde{\theta}^B\cr\bar{\theta}_{\bar{A}}\cr
\bar{\tilde{\theta}}_{\bar{B}}\cr}
}

\newsec{The fluctuation determinant}
The fermionic coordinates $\lambda$ act as Lagrange multipliers and
integrating over them leads to $\delta(\slash \hskip -6.5pt
D\theta)$. In other words the functional integral for the
fermionic coordinates $\theta$ is projected down onto their zero
modes. This is discussed in more detail in the context of light cone
string theory in \rtI . The Jacobian factor from this projection is
${\rm det}\bigl(\slash \hskip -6.5pt D^{\dagger}
\slash \hskip -6.5pt D\bigr)$.

In this section we focus on the contribution from the non zero modes
i.e. on the determinants.  For the four string scattering background
considered in this paper both the fermion term and the boson term
consist of four by four non-trivial blocks. Canceling part of the
trivial block of the bosonic determinant with the ghost determinant
the total fluctuation determinant $J$ reads
\eqn\fldet{
J={{\rm det}\bigl(\slash \hskip -6.5pt D^{\dagger}
\slash \hskip -6.5pt D\bigr)_{4\times 4}\over
{\rm det}\bigl(D^2)_{2\times 2}
{\rm det}^{1\over 2}
\bigl(D^2+{2i\over g_s}[F_{\rho\sigma},\,\,]\bigr)_{4\times 4}},
}
where  the indices
$\rho$, $\sigma$ take the values $0,9,1,2$, and
zero modes have been excluded from the determinants. The fermion
determinant is in four by four spinor space and the boson determinant
in four by four $\rho$, $\sigma$ space. 

It is well known that the fermionic and bosonic fluctuation
determinants
precisely cancel in a self dual background. Using the definition of
gamma matrices given in the appendix it is only a short calculation to
verify that this is indeed the case. Specifically for the instanton
background of section 2.2 we have
\eqn\ufu{
F_{\rho\sigma}=\pmatrix{
\matrix{F_{w\bar{w}}&D_wX^*\cr
-D_{\bar{w}}X& -F_{w\bar{w}}}
&0\cr
0&
\matrix{-F_{w\bar{w}}&D_{\bar{w}}X\cr
-D_wX^*&F_{w\bar{w}}}}
}
where we have we have used complex indices, i.e. rows labeled from top
to bottom by $w$,$x$,$\bar{w}$,$\bar{x}$ and columns labeled from left
to right by $\bar{w}$,$\bar{x}$,$w$,$x$. Using \fermL\ we have
\eqn\gf{
\slash \hskip -6.5pt D^{\dagger}\slash \hskip -6.5pt D
={1\over 2}(D^2+{2i\over g_s}[\tilde{F},\,\,])
\quad{\rm with}\quad
\tilde{F}=\pmatrix{
\hskip 5pt 0
&\matrix{\hskip 27pt & \hskip 27pt \cr \hskip 27pt & \hskip 27pt }\cr
\matrix{\hskip 27pt & \hskip 27pt \cr \hskip 27pt & \hskip 27pt }
&\matrix{F_{w\bar{w}} &D_wX^* \cr
-D_{\bar{w}}X &-F_{w\bar{w}}}},
}
leading to $J=1$. 

In other words if the background was self dual everywhere the
fluctuation determinant would be equal to one. The matrix string
background corresponding to four string scattering is however only
locally self dual around the branch points and its fluctuation
determinant would be corrected by perturbations from the non self dual
part away from the interaction points. The important point to note is
that, in the regime where it is justified to use a one loop
calculation, these are just small corrections and to lowest order can
be dropped. The conclusion from this analysis is that the interaction
points do not lead to a singular contribution to the determinant. This
is true even in the limit where $g_s$ and hence (via \lins ) 
the size of the instanton, tend to zero. 

It is interesting to compare this result with what one would expect
from a CFT fluctuation determinant around a branch point. In light
cone superstring theory
\rtI\ $\theta$ and $\lambda$ have the conformal
weights $1$ and $0$, respectively. Potentially there is a singular
contribution coming from the square root cut point which hides a world
sheet curvature singularity. Smoothing this out over some cut off
distance $\epsilon$ leads to a singular contribution
$1/\epsilon^{c\over 12}$ coming from the induced Liouville action for
the bosons and fermions. However for the choice of weights given above
the bosonic and fermionic central charges cancel and there is no
singular contribution. This can also be seen directly from a comparism
of the determinants (see \rtII ).

The conclusion of this section is that there are no singular
contributions to the determinant coming from the interaction
points and that this is consistent with the light cone superstring
calculation. This does not mean, however, that the fluctuation
determinant will be equal to precisely one. We will return to this
point in section 8. The remaining part of the functional integral is
an integration over a finite number of zero modes and collective
coordinates of the background configuration.  In the next two sections
we construct these modes.

\newsec{Instanton zero modes, local considerations}
In this section we focus on the field configurations in the neighbourhood 
of the instanton. The objective is to understand how the core of the
instanton, where the fields do not commute, effects the finiteness of
the zero mode field configurations. Far from the core of the instanton
the fields commute and the zero modes correspond to those defined on
a Riemann surface in the neighbourhood of a branch point.
In particular we will see that there are zero modes which, from the
Riemann surface point of view, would appear to diverge at the branch point,
but which are rendered finite by the non-commuting core of the
instanton. 
\vskip 20pt
\hskip 20pt\epsfbox{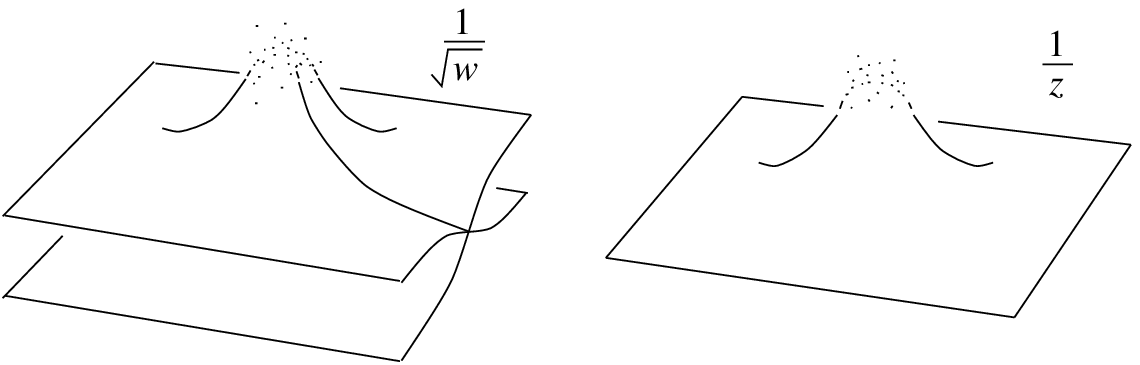}
\vskip 20pt
\centerline{{\bf Fig 2}. 
Finite field configurations for ``singular'' zero modes}
\centerline{shown in the $w$ plane and the $z=\sqrt{w}$ plane.}
\vskip 20pt
These ``singular'' modes will play a crucial role in the next section where
we construct global zero modes.
The instanton is embedded into a global classical solution
corresponding to a classical string world sheet.
The instanton zero modes will thus be embedded into global
zero modes which tend, asymptotically far down the strings, to constant
values. From the point of view
of the world sheet these modes can be written as holomorphic and/or 
anti-holomorphic functions of the cylindrical coordinates. The
singularities at the branch points are important since it is only with
singularities that one can construct non-trivial global modes.

The global zero modes will include the moduli for the classical
string world sheet along with their superpartners 
and zero modes for the abelian gauge theory defined on the string
world sheet. 

For the purposes of this section we will treat the instanton as being
embedded into the complex plane. Below we will explicitly construct
the bosonic and fermionic zero modes, imposing the condition that they
must be finite everywhere (except at infinity).

\subsec{Bosonic zero modes and collective coordinates}
The classical solution for the instanton \instsol\Gdef\difa\ 
involves only the two
dimensional gauge field $A$ and two of the 8 possible scalar fields,
$X_1$ and $X_2$.  There are three types of bosonic zero modes for the
fluctuations about this configuration characterized by which fields
they tend to asymptotically. Firstly there are the two-dimensional
gauge field zero modes. These come from the fact that the gauge fixing
term doesn't completely fix the gauge. In other words there are local gauge
transformations that are zero modes of the gauge fixing term.
The second correspond to translations
and deformations of the instanton solution, and asymptotically involve
only the scalar fields $X_1$ and $X_2$. Finally there are modes
involving $X_3,\cdots,X_8$. All three types are
zero modes of the eigenvalue equation for quadratic fluctuations about
the classical instanton background~:
\eqn\quadf{
\bigl[D_{\rho}D_{\rho}\eta_{\mu\nu}
+{2i\over g}[F_{\mu\nu},\,\,\,]\bigr]V^{(n)}_{\nu}
=\lambda^{(n)}V^{(n)}_{\nu}.
}
In this equation the $V^{(n)}_{\nu}$ are the fluctuation modes and
$\lambda^{(n)}$ their eigenvalues. For compactness the equation has
been written in its ten dimensional form. The ten dimensional indices 
$\mu,\nu=0,\cdots,9$ split into the two dimensional indices
$\alpha,\beta=0,9$, corresponding to the cylindrical coordinates, and
the indices $I=1,\cdots,8$ corresponding to the scalar matrix fields.
The covariant derivatives $D_{\mu}$ are covariant with respect to the
background fields, i.e. $D_{\mu}=\partial_{\mu}+i/g[A_{\mu},\,\,\,]$,
and $F_{\mu\nu}$ is the background field strength.
The standard  Feynman background gauge fixing term
$(D_{\mu}A_{\mu})^2$ has been used.

\subsec{Gauge field zero modes}
For the gauge field zero modes we search for gauge-fields, $V_{\mu}$,
that are locally pure gauge, 
\eqn\ggezer{
V_{\mu}=D_{\mu}\Lambda,
}
and which satisfy equation\quadf . Substituting \ggezer\  into
\quadf\  and commuting the covariant derivative to the right through
the $D^2$ term cancels the $F_{\mu\nu}$ term and leads us to look for
modes $\Lambda$ that satisfy
\eqn\gzm{
D_{\rho}D_{\rho}\Lambda=0,
}
i.e. to look for zero modes of the gauge fixing term. The resulting 
gauge fields $V_{\mu}$ must be finite everywhere (except
possibly at infinity) and must tend to some
function times $\tau_3$ at infinity so that they commute with the bosonic
fields. 
Note that equation\gzm\ is just the equation of motion for the 
complex bosonic field $X$. So we can immediately write down a
candidate zero mode~:
\eqn\lhalf{
\lambda^{({1\over 2})}=X^*.
}
with $X$ given by \XF . The gauge field zero mode 
$V^{(-{1\over 2})}=D_{w}\lambda^{({1\over 2})}$ is given by
\eqn\Vhalf{
V^{(-{1\over 2})}=
{1\over w^{1\over 2}}\bigl[
({1\over 2}\cosh\alpha+r\partial_r\alpha\,\sinh\alpha)\tau_3
+i({1\over 2}\sinh\alpha+r\partial_r\alpha\,\cosh\alpha)\tau_2
\bigr].
}
One still has to check that $V^{(-{1\over 2})}$
is finite at the origin, since it potentially diverges as $1/r$. Using
the fact that for small $r$ $\alpha(r)=-1/2\log r + a +{\cal O}(r^2)$,
we see that it is indeed finite at the origin~:
\eqn\Vhalfr{
V^{(-{1\over 2})}\rightarrow {r^{1\over 2}\over w^{1\over 2}}
{1\over 2}e^{-a}[\tau_3+i\tau_2]
\quad{\rm as}\quad
r\rightarrow\infty,
}
and is thus a valid zero mode. 

The most general solution for the gauge zero modes can be found 
by writing equation\gzm\ in terms of the two-dimensional fields and covariant
derivatives~:
\eqn\gzmtwo{
D_{\rho}D_{\rho}\Lambda=D_wD_{\bar{w}}\Lambda+D_{\bar{w}}D_w\Lambda
-{1\over
g^2}\bigl([X,[X^*,\Lambda]]+[X^*,[X,\Lambda]]\bigr)=0.
}
It is straightforward using \sfdl\ to see that it is satisfied by
\eqn\Lamlam{
\Lambda^{(n)}=\lambda^{(n)}+(\lambda^{(n)})^*,
}
with
\eqn\lzero{
\lambda^{(n)}=\cases{
w^{n-{1\over 2}}X^*\rightarrow w^n\tau_3
\quad{\rm for}\quad n={1\over 2},{3\over 2},{5\over 2},\cdots\cr
w^n 1_{2\times 2}
\quad\quad\quad{\rm for}\quad n=1,2,3,\cdots},
}
and
\eqn\Vzero{
V^{(n)}=D_w\lambda^{(n-1)}.
}
All these solutions are of
course written in the singular gauge in which asymptotically far from
the instanton all fields are diagonal. 

The physical meaning of these zero modes is given by their behaviour
far from the non-commuting core of the instanton where they can be
interpreted as zero modes of an abelian gauge field
defined on a Riemann surface in the neighbourhood of a branch
point. They are zero modes of the abelian gauge-fixing term
$(\partial_{\alpha}V_{\alpha})^2$. We see that we have the 
set of abelian pure gauge zero modes,
\eqn\Vw{
V_w=w^{-{1\over 2}},1,w^{1\over 2},w,w^{3\over 2},w^2,\cdots,
}
in the complex $w$ plane.  In particular there is a mode $V^{(-{1\over
2})}$ which is finite at the origin of the instanton but asymptotically
behaves as $w^{-{1\over 2}}$.  Unwinding the branch cut $w=z^2$,
$V_z=2\sqrt{w}V_w$ we see that we have the full set of abelian gauge
field zero modes,
\eqn\Vz{
V_z=1,z,z^2,z^3,z^4,\cdots,
}
defined in the complex $z$ plane. In section 4 we study the zero modes
about the classical solutions corresponding to light-cone Riemann
surfaces. In other words we embed the zero modes constructed above
into global abelian zero modes defined on the light cone Riemann
surfaces. The fact that we have the full set of zero modes \Vz\ means
that we will be able to construct arbitrary abelian zero modes on the
Riemann surface which are not in any way pinned to, or constrained by,
the interaction points.

The conclusion from this section is that for
the construction of the global zero modes for the complex gauge field 
$V$ we are allowed a 
$(w-w_0^+)^{-{1\over 2}}\sim
1/(z-z_0^+)$ 
singularity at $z=z_0^+$ and a 
$(w-w_0^-)^{-{1\over 2}}\sim
1/(z-z_0^-)$ singularity at 
$z=z_0^-$.

\subsec{Zero modes for the ghosts and for $X^3,\cdots,X^8$}
The zero mode equation for the ghosts is identical to equations
\gzm\gzmtwo . This allows to automatically write down all
their zero modes. They are given by the solutions for $\Lambda$
(equations \Lamlam\lzero ) of the previous section. 
The important point is that there is no mode that asymptotically
behaves as $w^{-{1\over2}}$ and is finite at the origin, so there will
be no non-trivial global zero mode we can construct.

Since the background only depends on the scalar fields
$X^1$ and $X^2$ the zero mode equation for the scalar fields
$X^3,\cdots,X^8$, contains no $F_{\mu\nu}$ term and is thus also identical
to equations\gzm\gzmtwo . So again there will be no non-trivial zero modes.

\subsec{Zero modes for $X^1$ and $X^2$}
The most obvious zero mode to write down for the fields
$X=1/\sqrt{2}(X^1+iX^2)$ and $X^*$ is that corresponding to the
translation of the instanton configuration. To construct this mode one
proceeds as for the four dimensional instanton by taking
the derivative of all the fields and at the same time shifting them by a
gauge transformation so that the background gauge fixing condition,
$D_{\mu}V_{\mu}$, is preserved. The result being that the zero modes of
\quadf\ that correspond to translations are given by
\eqn\trzm{
V_{\mu}=\partial_{\alpha}A_{\mu}-D_{\mu}A_{\alpha}=F_{\alpha\mu}.
}
The first term corresponds to translation in the $\alpha$ direction,
the second term is the gauge transformation with gauge parameter
$A_{\alpha}$. Substituting \trzm\ into \quadf\ and using the equations
of motion and the Bianchi identity this is indeed seen to be a zero
mode of \quadf . Written in terms of the two dimensional complex
fields of section 2 the solution reads
\eqn\trzmtwo{
\pmatrix{V\cr Y}=
\pmatrix{-F_{w\bar{w}}\cr D_{\bar{w}}X}\quad\longrightarrow\quad
\tau_3\pmatrix{0\cr \bar{w}^{-1/2}}.
}
Its asymptotic behaviour, indicated on the r.h.s, shows that from the
world sheet point of view it is singular at the origin. The
non-commutative core of the instanton, however, renders the field
configuration finite at the origin (see equation\Vhalfr ). As already
stressed the fact that it is a singular zero mode from the world sheet
point of view will allow us to construct a non-trivial global zero
mode.  It will correspond to deforming the classical world sheet by
moving a branch point.

It is important to check to see if their are any other singular zero
modes. The only other possibility would be a mode asymptotically
behaving as $w^{-{1\over 2}}$. We write
equation \quadf\ in terms of the two dimensional covariant
derivatives, field strengths and scalar fields~:
\eqn\bszm{
\Biggl\{ 
D_{\rho}D_{\rho}
+2{i\over g}\pmatrix{[F_{w\bar{w}},\,\,\,]&[D_wX^*,\,\,\,]\cr
           [D_{\bar{w}}X,\,\,\,]&-[F_{w\bar{w}},\,\,\,]}\Biggr\}
\,\,\,
\pmatrix{V\cr Y}=0,
}
where $D_{\rho}D_{\rho}$, in terms of two dimensional covariant
derivatives and fields, has already been given in \gzmtwo . It is
possible to find simple generalizations of \trzmtwo , satisfying
\bszm , for arbitrary half-integer powers of $\bar{w}$~:
\eqn\bzm{
\pmatrix{V\cr Y}=
\pmatrix{
-\bar{w}^nF_{w\bar{w}}\cr 
D_{\bar{w}}(\bar{w}^nX)} 
\quad\longrightarrow\quad
\tau_3\pmatrix{0\cr \bar{w}^{n-1/2}}
\quad\quad n\geq 0
}
along with the trivial solution
\eqn\bzmint{
Y=\bar{w}^n 1_{2x2}\quad\quad n\geq 0.
}

However there is no obvious, simple generalization of \bzm\ for powers
of $w$. It would seem
natural, however, that there would NOT be a mode behaving as
$w^{-{1\over 2}}$ which is also finite at the origin. 

We will thus assume that this is the case (to prove it would mean
analyzing carefully the coupled differential equations of \bszm ).
With this assumption the conclusion from this section then is that for
the construction of the global zero modes we are allowed a 
$(\bar{w}-\bar{w_0^+})^{-{1\over 2}}\sim
1/(\bar{z}-\bar{z}_0^+)$ 
singularity at $z=z_0^+$ and a 
$(w-w_0^-)^{-{1\over 2}}\sim
1/(z-z_0^-)$ singularity at 
$z=z_0^-$.

\subsec{Fermionic zero modes}
The fermion zero mode equation reads
\eqn\fmodes{
\slash\hskip-7ptD\theta=0,
}
with $\slash\hskip-7ptD\theta$ given explicitly in \fermL .  The most
important zero mode is that coming from the Euclidean version of the
infinitesimal Minkowski space supersymmetry transformation,
\eqn\fsyzm{
\theta=F_{\mu\nu}\Gamma_{\mu\nu}\epsilon,
}
where $\epsilon$ is a constant ten-dimensional Majorana-Weyl spinor.
This can be show to be a solution of \fmodes\ by using the gamma
matrix identity 
$\Gamma_{\rho}\Gamma_{\mu\nu}
=\Gamma_{\rho\mu\nu}
+1/2(\eta_{\rho\mu}\Gamma_{\nu}-\eta_{\rho\nu}\Gamma_{\mu})$,
followed by the Bianchi identity and the equation of motion for
$F_{\mu\nu}$. 

It is easy to see from \fermL\ that the Euclidean fermion zero mode
equivalent to \fsyzm\ has the form
$\theta=\tilde{\theta}=0$ and
\eqn\tffzm{
\pmatrix{\bar{\theta}\cr\bar{\tilde{\theta}}}=
\pmatrix{-F_{w\bar{w}} &D_{\bar{w}}X \cr
D_wX^* &-F_{w\bar{w}}}
\pmatrix{\bar{\tilde{\theta}}^p\cr\bar{\theta}^p}
\rightarrow
\pmatrix{\bar{\theta}^p\bar{w}^{-{1\over 2}}\cr
\bar{\tilde{\theta}}^p w^{-{1\over 2}}},
}
where $\bar{\theta}^p$ and $\bar{\tilde{\theta}}^p$ are two component 
constant, complex spinor variables. 
The asymptotic behaviour on the r.h.s. of \tffzm\ shows that from the
world sheet point of view the mode is singular at the origin. As for 
the bosonic zero modes the field configuration is in fact finite at the
origin.

Again it is important to check that there are no other ``singular''
zero modes. Using the explicit expression for $\slash \hskip -6.5pt D$
of \fermL\ it is relatively simple to see that $\theta$ and
$\tilde{\theta}$ have the same solutions, respectively, as
$\bar{\lambda}^{(n)}$ and $\lambda^{(n)}$ (equations \gzmtwo\Lamlam ).
In other words there is no zero mode for $\theta$ or $\tilde{\theta}$ that
from the world sheet point of view appears to have a singularity at
the branch point but is nevertheless finite at the origin. 

For $\bar{\theta}$ and $\bar{\tilde{\theta}}$ we have already found the 
``singular''
zero mode \fsyzm\ and there are no other possibilities (we cannot, for
example, have $\bar{\theta}$ behaving as $w^{-{1\over 2}}$ since
asymptotically $\bar{\theta}$ must be an anti-holomorphic function of
$w$). 

The fermion zero mode \tffzm\ is for the self-dual configuration
around the branch point $w=w_0^+$. The zero mode around the
anti-self-dual configuration\footnote{$^3$}{For the anti-self-dual
instanton we have $D_wX^*=0$, $D_{\bar{w}}X=0$ and
$F_{w\bar{w}}+{i\over g_s}[X^*,X]=0$} 
of the $w_0^-$ branch point can also be
read off from \fermL . It is given by $\bar{\theta}=\bar{\tilde{\theta}}=0$ 
and
\eqn\antitffzm{
\pmatrix{\theta\cr\tilde{\theta}}
=\pmatrix{F_{w\bar{w}}&D_{\bar{w}}X^*\cr
D_wX&F_{w\bar{w}}}
\pmatrix{\tilde{\theta}^p\cr\theta^p}\rightarrow
\pmatrix{\theta^p \bar{w}^{-{1\over 2}}\cr\tilde{\theta}^p w^{-{1\over 2}}}.
}

The conclusion is that for the construction of global fermion zero
modes in the next section $\theta$,$\tilde{\theta}$,$\bar{\theta}$ and
$\bar{\tilde{\theta}}$ can respectively have the singularities 
$1/(\bar{z}-\bar{z}_0^-)$,
$1/(z-z_0^-)$,
$1/(\bar{z}-\bar{z}_0^+)$,
and $1/(z-z_0^+)$.

\newsec{Zero modes, global considerations}
In the previous section we analyzed in detail the bosonic and
fermionic zero modes in the neighbourhood of the instanton
configuration for the branch points of the matrix string world sheet.
In particular we identified zero mode field configurations which are
finite at the branch points but from the world sheet point of view
appear to have a singularity at the branch point. In this section we
take these zero modes and embed them into non-trivial zero modes for the
gauge, boson and fermion fields for global classical solutions
corresponding to light cone Riemann surfaces.

We will construct global zero modes which tend to constant values
asymptotically far down the strings.  
Specifically we will construct modes satisfying the boundary conditions
\eqn\gbfbc{
V_{\tau}=\partial_{\tau}V_\sigma=0,
\quad
\partial_{\tau}Y^I=0,
\quad
\partial_{\tau}\theta=0
\quad\quad{\rm for}\quad\quad
\tau=\tau_{\rm i}\,\,{\rm or}\,\,\tau_{\rm f},
}
where $\tau_{\rm i}$ and $\tau_{\rm f}$ are the initial and final
times.

From the world sheet point of view all zero modes can be decomposed
as holomorphic and/or anti-holomorphic functions of $z$, with
possible simple pole singularities at the branch points. At the end of
sections 4.2,4.3,4.4 and 4.5 we enumerated the possible pole
structures for the bosonic and fermionic modes. In the following
sections we use these singularities to explicitly construct non-trivial
global zero modes satisfying the boundary conditions \gbfbc .

\subsec{Abelian gauge and ghost field zero modes}

The two dimensional gauge field $V=1/\sqrt{2}(V_{\tau}-iV_{\sigma})$ 
must be a holomorphic function of $w$ or
equivalently of $z$. We have already seen in the previous section that
it can have the poles 
$1/(z-z_0^+)$
at the branch point $z_0^+$ and 
$1/(z-z_0^-)$
at the branch point $z_0^-$. This allows
us to construct the following zero mode
\eqn\glgzm{
V=i{\sum_i\epsilon_i{a_i\over z-z_i}\over
\sum_i\epsilon_i{N_i\over z-z_i}}\rightarrow{a_i\over N_i}
\quad{\rm for}\quad z\rightarrow z_i,
}
with the $a_i$ real numbers satisfying
\eqn\aicd{
\sum_i\epsilon_ia_i=0.
}
Note from equation \rts\ that the numerator of \glgzm\ has zeros 
at the branch points and hence gives the correct pole structure

Asymptotically far down a string
($z\rightarrow z_i$) the boundary conditions \gbfbc\ are satisfied with
$V_{\sigma}$ tending to the constant value $a_i/N_i$ as indicated in
\glgzm . A Wilson loop encircling the string thus generates the 
$U(1)$ element $e^{{i\over g_s}a_i}$.

The mode \glgzm\ is the most general abelian differential defined
on the string diagram Riemann surface.

For the ghost field zero modes no singularities are allowed at the
branch points and there is thus just a single, globally constant zero
mode.

\subsec{Bosonic field zero modes}
The non-trivial bosonic zero modes satisfying the boundary conditions
\gbfbc\ 
involve only
the bosonic field $X=1/\sqrt{2}(X_1+iX_2)$ and it's complex
conjugate. 
From the analysis of
section 4.4 the zero modes can have 
can have a
$1/(\bar{z}-\bar{z}_0^+)$ 
pole at $z=z_0^+$ and a 
$1/(z-z_0^-)$ pole at 
$z=z_0^-$.

The most general zero mode,
with such singularities, satisfying \gbfbc\ is given by a simple
generalization of \glgzm ~:
\eqn\glbzm{
Y=-b{\sum_i\epsilon_i{p_i\over z-z_i}\over
\sum_i\epsilon_i{N_i\over z-z_i}}
-\tilde{b}^*
{\sum_i\epsilon_i{p_i\over \bar{z}-\bar{z}_i}\over
\sum_i\epsilon_i{N_i\over \bar{z}-\bar{z}_i}}
\quad\rightarrow\quad-(b+\tilde{b}^*){p_i\over N_i}
\quad{\rm for}\quad z\rightarrow z_i,
}
where the $b$ and the $\tilde{b}$ are complex numbers.
The numerators of \glbzm\ have zeros at $z=z_0^+$ and $z=z_0^-$ (see
equations \rts ). The denominators of the holomorphic and
anti-holomorphic parts have zeros at respectively $z=z_0^+$ and 
$\bar{z}=\bar{z}_0^-=z_0^+$ (see equations \holbpt\antiholbpt ). So in
total the zero mode \glbzm\ has the correct pole structure. Choosing
the standard positions for three of the incoming and outgoing strings
$z_1=0$, $z_2=1$, $z_4=\infty$ and $z_3=\lambda$ we can write the zero
mode as the simpler expression
\eqn\glbzmtr{
Y=-(b+\tilde{b}^*){p_4\over N_4}+
\Bigl({p_1\over N_1}-{p_4\over N_4}\Bigr)
\Biggl(b{z_0^-\over z-z_0^-}
+\tilde{b}^*{\bar{z}_0^+\over\bar{z}-\bar{z}_0^-}\Biggr).
}
We could, of course, have written down the overall form of this mode directly
from the constraints on its singularities specified above. The advantage,
of the more complicated form \glbzm\ is that we can read off
immediately its behaviour asymptotically far down the
strings which we indicate on the r.h.s of \glbzm .

The holomorphic term in \glbzmtr\ corresponds to translation of the
branch point at $z_0^-$ and the anti-holomorphic term to the
translation of the branch point $z_0^+$. To see this more explicitly
note that in the neighbourhood of one of the $z_0$'s we can write
\eqn\bpttr{
w=w_0+{1\over 2}\alpha(z-z_0)^2+\cdots
\quad{\rm and}\quad
X=X(z_0)+\beta(z-z_0)+\cdots
}
Inverting the first equation to write $z-z_0=\sqrt{2/\alpha(w-w_0)}$
and substituting into the second we arrive at an expression for $X$ in
terms of $w$. We can then differentiate with respect to the branch
point position $w_0$ to find
\eqn\dwX{
\partial_{w_0}X={\beta\over\alpha}{1\over z-z_0^-}\,+\,\cdots.
}
with $\alpha$ and $\beta$ given by
\eqn\addwbdX{
\alpha=\partial_z^2w|_{z=z_0}
\quad{\rm and}\quad
\beta=\partial_zX|_{z=z_0}.
}
Evaluating $\alpha$ and $\beta$ explicitly one finds
\eqn\dwXII{
\partial_{w_0^-}X=\Bigl({p_1\over N_1}-{p_4\over N_4}\Bigr)
{z_0^-\over z-z_0^-}\,+\,\cdots
\quad{\rm and}\quad
\partial_{\bar{w}_0^+}X=\Bigl({p_1\over N_1}-{p_4\over N_4}\Bigr)
{\bar{z}_0^+\over \bar{z}-\bar{z}_0^+}\,+\,\cdots.
}
The final outcome of this analysis being that one can identify the
shift in the branch points $\delta w_0^+$ and $\delta w_0^-$ with the
parameters $b$ and $\tilde{b}$ via
\eqn\dwbdwb{
b=\delta w_0^-
\quad{\rm and}\quad
\tilde{b}=\delta w_0^-.
}

Out of the four parameter space two parameters translate the
whole classical solution in the $\tau$ and
$\sigma$ directions, and two change the relative $\tau$ and $\sigma$
separations of the branch points.

Finally, in addition to the non-trivial bosonic zero modes constructed
above, there are eight constant modes corresponding to globally
translating the world sheet in the eight dimensional transverse target
space.

\subsec{Fermion zero modes}

From section 4.5 the pole structures allowed for the fermion 
zero modes are 
a $1/(\bar{z}-\bar{z}_0^-)$ pole for $\theta$, 
a $1/(z-z_0^-)$ pole for $\tilde{\theta}$,
a $1/(\bar{z}-\bar{z}_0^+)$ pole for $\bar{\theta}$ and
a $1/(z-z_0^+)$ for $\bar{\tilde{\theta}}$.

The construction of the fermion zero modes is thus virtually identical 
to that for the bosons; they are given by
\eqn\tho{
\theta=
{\sum_i{\epsilon_i p_i^*\over \bar{z}-\bar{z}_i}\over
\sum_i{\epsilon_i N_i\over \bar{z}-\bar{z}_i}}\theta^p
=\bigl({p_4^*\over N_4}-\bigr({p_1^*\over N_1}-{p_4^*\over N_4}\bigr)
{\bar{z}_0^-\over \bar{z}-\bar{z}_0^-}\bigr)\theta^p\quad
\rightarrow\quad {p_i^*\over N_i}\theta^p,
}
\eqn\thtw{
\tilde{\theta}=
{\sum_i{\epsilon_i p_i\over z-z_i}\over
\sum_i{\epsilon_i N_i\over z-z_i}}\tilde{\theta}^p
=\bigl({p_4\over N_4}-\bigr({p_1\over N_1}-{p_4\over N_4}\bigr)
{z_0^-\over z-z_0^-}\bigr)\tilde{\theta}^p\quad
\rightarrow\quad {p_i\over N_i}\tilde{\theta}^p,
}
\eqn\thth{
\bar{\theta}=
{\sum_i{\epsilon_i p_i\over \bar{z}-\bar{z}_i}\over
\sum_i{\epsilon_i N_i\over \bar{z}-\bar{z}_i}}\bar{\theta}^p
=\bigl({p_4\over N_4}-\bigr({p_1\over N_1}-{p_4\over N_4}\bigr)
{\bar{z}_0^+\over \bar{z}-\bar{z}_0^+}\bigr)\bar{\theta}^p\quad
\rightarrow\quad {p_i\over N_i}\bar{\theta}^p,
}
\eqn\thfo{
\bar{\tilde{\theta}}=
{\sum_i{\epsilon_i p_i^*\over \bar{z}-\bar{z}_i}\over
\sum_i{\epsilon_i N_i\over \bar{z}-\bar{z}_i}}\,\bar{\tilde{\theta}}^p
=\bigl({p_4^*\over N_4}-\bigr({p_1^*\over N_1}-{p_4^*\over N_4}\bigr)
{z_0^+\over z-z_0^+}\bigr)\bar{\tilde{\theta}}^p\quad
\rightarrow\quad {p_i^*\over N_i}\bar{\tilde{\theta}}^p,
}
where $\theta^p$,$\tilde{\theta}^p$,$\bar{\theta}^p$ and 
$\bar{\tilde{\theta}}^p$ are constant two component
complex spinor variables. The simpler expressions are written for
the standard choice $z_1=0$,$z_2=1$,$z_4=\infty$ and $z_3=\lambda$. 
Asymptotically far down a string
($z\rightarrow z_i$) the fermion zero modes tend to the constant
values indicated on the r.h.s. of the above expressions. As we will 
see in section 8 these asymptotic values determine the kinematic 
structure of the scattering amplitudes. 

In addition to the non-trivial fermion zero modes constructed 
above, there are eight, globally constant fermion modes 
$\theta^N$,$\bar{\theta}^N$,$\tilde{\theta}^N$ and $\bar{\tilde{\theta}}^N$
which take the same constant value on each of the four strings.

\newsec{Fermion zero modes and graviton wavefunctions}
In this section we will focus on the fermionic zero modes for a single
string. This is equivalent to studying the zero modes of the previous
section asymptotically far down one of the strings. We will construct
ground state wave functions from these modes. 

The decomposition of the fermionic variables \defth\ into coordinates
and momenta explicitly breaks the $spin(8)$ invariance down to
$U(1)\otimes U(1)\otimes SU(2) \otimes SU(2)$. Using the
transformation properties of the $\theta$ under $SO(8)$ rotations
however one can build up from combinations of the $\theta$ the
components of the ${\bf 8}_s$,${\bf 8}_c$ and ${\bf 8}_v$
representations of $SO(8)$. We will restrict our attention here to the
vector representation ${\bf 8}_v$ and will construct one set of vector
components from $\theta^A$ and $\bar{\theta}_{\bar{A}}$ for the left
moving sector and another set of vector components from 
$\tilde{\theta}^A$ and $\bar{\tilde{\theta}}_{\bar{A}}$ for the right
moving sector. Tensored together they will then form the ground
state quantum numbers for the $SO(8)$ polarizations of the graviton
$G^{IJ}$, antisymmetric tensor field $B^{IJ}$ and the dilaton $\Phi$.
Specifically, with polarization tensor $\epsilon^{IJ}$ the wavefunction
reads
\eqn\wvfnc{
\Psi(\epsilon^{IJ},\theta,\bar{\theta},\tilde{\theta},\bar{\tilde{\theta}})
=\Psi^I(\theta,\bar{\theta})
\,\tilde{\Psi}^J(\tilde{\theta},\bar{\tilde{\theta}})
\,\epsilon^{IJ},
}
with $\Psi$ and $\tilde{\Psi}$ denoting respectively the left and
right moving parts of the wavefunction
We will just focus on the left moving sector, since (before
integration over the moduli) the left and
right sectors are completely decoupled and the analysis for the right
movers is identical to that for the left movers. As a
book keeping device we will contract the left moving components with a
polarization vector $\xi$. Using the $SO(8)$ transformation properties of 
$\theta^A$ and $\bar{\theta}_{\bar{A}}$ enumerated in section 3 and
the appendix the $SO(8)$ vector separates into three parts
corresponding to the subspaces $(X^1,X^2)$, $(X^3,X^4)$ and
$(X^5,X^6,X^7,X^8)$
\eqn\Lwvfnc{
\eqalign{
\Psi(\xi^I,\theta,\bar{\theta})=&\Psi^I(\theta,\bar{\theta})\xi^I\cr
=&\Psi_{(12)}(\xi,\xi^*,\theta,\bar{\theta})
+\Psi_{(34)}(\tilde{\xi},\tilde{\xi}^*,\theta,\bar{\theta})
+\Psi_{(5678)}(\xi^m,\theta,\bar{\theta}),
}
}
where the indices indicate the corresponding subspaces.
The $\Psi_{(12)}$, $\Psi_{(34)}$ and $\Psi_{(5678)}$ are given by
\eqn\ottffsse{
\eqalign{
&\Psi_{(12)}(\xi,\xi^*,\theta,\bar{\theta})
=\,\,ip^+(\xi^*\theta^2+\xi\bar{\theta}^2)\cr
&\Psi_{(34)}(\tilde{\xi},\tilde{\xi}^*,\theta,\bar{\theta})
=\,\,\tilde{\xi}+(p^+)^2\tilde{\xi}^*\theta^2\bar{\theta}^2 \cr
&\Psi_{(5678)}(\xi^m,\theta,\bar{\theta})
=\,\,p^+\xi^m\theta\sigma^m\bar{\theta},\cr
}
\quad\quad{\rm with}\quad\quad
\eqalign{
\xi,\xi^*=&{1\over 2}(\xi^1\pm i\xi^2)\cr
\tilde{\xi},\tilde{\xi}^*=&{1\over 2}(\xi^3\pm i\xi^4)\cr
}
}
The standard definitions for $\theta^2$, $\bar{\theta}^2$ and
$\theta\sigma^m\bar{\theta}$ are given in the appendix.  Note that to
obtain the correct $U(1)$ charges for the components of the
polarization vector $\xi$ the wavefunction $\Psi$ must be assigned the
$U(1)$ charges $(0,1)$ with respect to rotations in respectively the
$X^1$,$X^2$ and $X^3$,$X^4$ planes. Each $\theta$ is weighted with a
factor of $\sqrt{p^+}$, where $p^+$ is measured incoming to the
process. In other words for the two strings $1$ and $2$ $p^+$ is
replaced respectively by $N_1$ and $N_2$ whereas for the two outgoing
strings $3$ and $4$ $p^+$ is replaced by $-N_3$ and $-N_4$. The powers
of $p^+$ are needed to
reproduce Lorentz invariant results.\footnote{$^4$}{We deduce the
necessary factors, by observing that the resulting scattering
amplitude is Lorentz invariant. A more rigorous way of determining
them, however, would be to construct the $SO(8)$ rotation generators
out of the $\theta$ and $\lambda$ and verify that, with the relative
factors of \ottffsse , $\Psi^I$ is indeed an $SO(8)$ vector. We do 
not pursue this here.}

Finally note that in light cone gauge an arbitrary physical
polarization vector $\xi$ can be constructed by specifying
only its components in the transverse space. The equations of motion 
impose the constraint
\eqn\polcstr{
k.\xi=0.}
There is a remaining gauge degree of freedom which leaves 
the physical polarization condition \polcstr\ unchanged and
corresponds to adding a multiple of momenta $k$ to $\xi$.
By gauge choice one can set the $+$ component of the polarization to
zero. We thus have the physical polarization given by
\eqn\physpol{
\xi=(\xi^+,\xi^-,\xi,\xi^*,\tilde{\xi},\tilde{\xi}^*,\xi^m),}
with
\eqn\xipxim{
\xi^+=0
\quad\quad{\rm and}\quad\quad
\xi^-={1\over N}(\xi p^*+\xi^* p).}
Likewise an arbitrary physical polarization tensor $\epsilon$ can be
specified by its transverse components.

\newsec{The four graviton scattering amplitude}
The calculation involves integrating over all modes. The non zero
modes lead to determinants, whereas the zero modes must be soaked up
into the wavefunctions, or, in the case of the gauge field zero mode,
will be integrated over a finite interval. In this section we study in
detail the integration over the zero modes. At the end we return to
the fluctuation determinant and combine the results to obtain the four
graviton scattering amplitude.

\subsec{Integration over bosonic zero modes}
In section 6.2 we constructed four non-trivial zero modes for the
bosonic fields.  In addition to these modes there are eight constant
bosonic zero modes corresponding to shifting the whole string
configuration through the eight dimensional transverse
space. Integration over these zero modes leads to transverse momentum
conservation.

The non-trivial bosonic zero modes of \glbzm\glbzmtr\ correspond to
translations of the branchpoints. Two of the possible four modes,
$b=\tilde{b}=\delta\tau$ and $b=\tilde{b}=i\sigma$ correspond to
translation of the whole classical solution in the $\tau$ and $\sigma$
directions respectively. Integration over the first of these modes
leads to conservation of the light cone energy, $p^-$. Integration
over the second leads to invariance under shift of $\sigma$

The two others correspond to the relative displacement of the branch
points.  The integral over these modes is not suppressed by a gaussian
term in the action. We thus have to integrate over them carefully.

Let us denote by $w_r=\tau_r+i\sigma_r$ the point
in moduli space of the four string scattering diagram. It specifies
the relative displacement of the branchpoints with respect to each
other. The point $w_r=0$ is the saddle point world sheet. 
For small $w_r$ we can identify $w_r$ with the parameters $b$ and
$\tilde{b}$ of \dwbdwb\ i.e. $w_r=b=-\tilde{b}$. 
Denote $X_c(w,w_r)$ the classical matrix string field configuration
corresponding to the point $w_r$. Varying $w_r$ from $w_r$ to $w_r+\delta w_r$
takes us from one classical configuration to another. The change of
fields corresponding to this deformation are, by the equations of
motion, zero modes of the quadratic operator; 
\eqn\ytys{
Y_{\tau}(w,w_r)=\partial_{\tau_r}X_c(w,w_r)
\quad\quad{\rm and}\quad\quad
Y_{\sigma}(w,w_r)=\partial_{\sigma_r}X_c(w,w_r).}
However, zero modes which tend to a non-zero constant value
asymptotically far down a string will not decouple from the
action. The background field configuration \classS\ is linear in
$\tau$ and in integrating the kinetic term $(\partial (X_c+Y))^2$ by
parts one picks up a boundary term, which couples the zero mode
linearly to the classical background. For the saddle point world sheet
configuration the zero mode corresponding to relative displacement of
the branch points tends to zero asymptotically down the strings, and
hence does decouple. This is another way of seeing that the background
is a stationary point of the action. It also indicates that one needs to switch
from integrating over this mode to integrating over the modular
parameter $w_r$.

A general field configuration can be decomposed in two ways. We can
set $w_r=0$, i.e. start with the saddle point classical configuration, and
expand using a complete set of modes about this background. 
Alternatively we can
keep $w_r$ as a free complex variable and expand in the set of modes
excluding the zero modes
$Y_{\tau}$ and $Y_{\sigma}$ of \ytys . We thus have the identity
\eqn\udecomp{\eqalign{
X(w)=&X_c(w,0)+\sum_ny_nY_n(w,0)\cr
=&X_c(w,w_r)+\sum_{n\neq \tau\sigma}\tilde{y}_nY_n(w,0).}
}
We know the integration measure for the first decomposition. It is
determined by the condition.
\eqn\dXmescond{
\int(dY)e^{-\pi<Y|Y>}=1
\quad{\rm with}\quad
<Y|Y'>={1\over 2\pi^2}\int d^2w \Tr[\bar{Y}(w)Y'(w)+\bar{Y}'(w)Y(w)].
}
Substituting in the first decomposition of \udecomp\ we find the
zero mode measure
\eqn\dXmes{
\int(dY)_0=\int\prod_q dy_q
\bigl|{\rm det}<Y_q|Y_r>\bigr|^{1\over 2}.}

To obtain the integration measure for the second decomposition we need
to calculate the jacobian for passing from integration variables
$x_{\tau}$,$x_{\sigma}$ and $x_n$
to integration variables $\tau_r$,$\sigma_r$ and $\tilde{x}_n$, 
$(n\neq \tau,\sigma)$. 

From \udecomp\ we have the identity
\eqn\umd{
\sum_ny_nY_n(w)=X(w,w_r)-X(w,0)+\sum_{n\neq\tau,\sigma}\tilde{y}_nY_n(w),
}
where the sums are over all modes.
Taking inner products we can relate the two sets of variables
\eqn\mdmd{
\eqalign{
&y_{\tau}={<Y_{\tau}|X(w,w_r)-X(w,0)>\over <Y_{\tau}|Y_{\tau}>}\cr
&y_{\sigma}={<Y_{\sigma}|X(w,w_r)-X(w,0)>\over <Y_{\sigma}|Y_{\sigma}>}\cr
&y_n={<Y_n|X(w,w_r)-X(w,0)>\over <Y_n|Y_n>}\,\,+\tilde{y}_n
\quad{\rm for}\quad n\neq\tau,\sigma\cr
}
}
The jacobian for passing from one set of variables to the other
is seen to be given by the finite dimensional determinant
\eqn\jac{
{\cal J}={\rm det}
\pmatrix{\partial_{\tau_r} y_{\tau}&\partial_{\tau_r} y_{\sigma}\cr
         \partial_{\sigma_r} y_{\tau}&\partial_{\sigma_r} y_{\sigma}}
\,\,\sim\,\,1,
}
where we have used \umd\ytys\ to see that, to lowest order, ${\cal J}=1$. 
Corrections to this approximation form a power series in $\tau$
and $\sigma$. In the domain of validity of the one loop approximation,
where the momenta $p$ are very large, these terms can be dropped since
the integrals over $\tau$ and $\sigma$ are entirely dominated by their
quadratic corrections to the minimal action.
The conclusion is that the integration measure for the collective
coordinates $\tau$ and $\sigma$ and for the other zero modes is
exactly as for \dXmes\ except with $dy_{\tau}$ replaced by $d\tau_r$ and
$dy_{\sigma}$ replaced by $d\sigma_r$~:
\eqn\colzmdmes{
\int(dY)_0=\int d\tau_r d\sigma_r
\prod_{q\neq\tau,\sigma} dy_q
\bigl|{\rm det}<Y_q|Y_r>\bigr|^{1\over 2}.
}

By the choice of the decomposition \udecomp\ the only dependence of
the amplitude on the relative displacements $\tau_r$ and $\sigma_r$ is
through the classical action. Expanding the classical action up to
quadratic order in $\tau_r$ and $\sigma_r$ the integrations over
$\tau_r$ and $\sigma_r$ can then be performed. The integrals have to
be evaluated by the method of steepest descent (see \gmII) 
leading to a factor of $i$, the final result being
\eqn\wdrint{
\int d\tau_rd\sigma_r
\,\,e^{S(w_r)}=i\,\,{c\over sut}\,\,
e^{-{1\over 4}(s\log s+t\log t+u\log u)},}
with $s$,$t$ and $u$ the Lorentz invariants defined in \stu\stuexp .
The non-Lorentz invariant factor $c$ is given, up to numerical factors, by 
\eqn\defc{
c=s^2N_4^2|z_0^+-z_0^-|^2,}
where $z_0^+$ and $z_0^-$ are the two branch point positions in the
complex $z$ plane. They are determined by the quadratic equation \rts\
written with the standard choice $z_1=0$, $z_2=1$, $z_3=\lambda$,
$z_4=\infty$, with $\lambda$ set to the value corresponding to a
stationary point of the classical action (see \ghv\ for details).

The non-Lorentz invariant factor $c$ might at first
appear problematic. However precisely this factor 
occurs in a light-cone gauge calculation for the same amplitude in 
superstring theory or bosonic string theory where it is well known 
that light-cone gauge produces Lorentz invariant results. It is
thus instructive to see how Lorentz invariance is restored in the case of
light cone string theory. Described in detail in chapter 11 (see in
particular appendix 11A) of Green Schwarz Witten \gsw\ is a careful
evaluation of the fluctuation determinant (involving a regularization
of the branch points) using the trace anomaly. The result being that
the determinant contains precisely the right Jacobean factor to convert
the integral $\int d\tau_r d\sigma_r$ into the integral 
$\int d^2\lambda$ where $\lambda$ is the standard complex moduli for
the four string scattering process already introduced in the previous
paragraph. In terms of the complex moduli $\lambda$ the classical
action \classSmink\ reads
\eqn\classSminklambda{
S = k_1.k_3\log|\lambda|+k_2.k_3\log|1-\lambda|.
}
Integration over $\lambda$ clearly leads to a Lorentz invariant
result.

In the light of the above discussion it seems plausible that the
matrix string determinant for the amplitude studied in this paper
could likewise generate the same Jacobean factor. To show this
explicitly, however, would involve finding some generalization of the
the trace anomaly techniques applicable to the non-abelian determinant
of \fldet . We do not address this question more fully here.

\subsec{Integration over gauge field and ghost zero modes}
There are three nontrivial gauge boson modes \glgzm\ , corresponding
to the fact that a generic gauge field zero mode takes on four
different values, $a_i\over N_i$, down the four strings, subject to
the constraint $\sum_i \epsilon_ia_i=0$.

As for the bosonic zero modes of the previous
section their integration measure includes a determinant of their
inner products.
\eqn\gbmes{
\int (dV)_0=\prod_{i=1}^4\int_0^{2\pi g_s}da_i\,\,\,
\bigl|{\rm det}<V_i|V_j>\bigr|^{1\over 2}\,\,\,\delta(a_1+a_2-a_3-a_4),
}
where the determinant is a
three by three determinant and can be calculated by choosing any
linearly independent basis for the $V_i$. One such basis would be to
say that $V_i$ $(i=1,2,3)$ is the mode which tends to $1$ down the
$i$th string, tends to $\pm 1$ down the fourth string and tends to
zero down the others.  The integrals only go up to $a_i=g_s$
since there is a globally well defined gauge transformation that
identifies $a_i=g_s$ with $a_i=0$. Specifically for the string
$i$ the globally defined unitary matrix is $U=e^{-i\sigma/N_i}$.

Integrating over the gauge field zero modes gives rise to the
contribution
\eqn\gzmct{
(2\pi g_s)^3\,\,\bigl|{\rm det}<V_i|V_j>\bigr|^{1\over 2}.
}

There is a single ghost field zero mode, the constant mode. 
It comes from the fact that there is a globally constant $U(1)$ gauge
transformation. To understand how to deal with this mode, let us look
carefully at the Fadeev Popov representation of the gauge fixing
determinant. We insert into the functional integral the
factor of $1$ written as
\eqn\fp{
1=\Delta\,\,\int(dU)\delta'(f-D_{\mu}V_{\mu}^{(U)}),}
where $\int(dU)$ is the Haar measure for the integral over the unitary
group and $V_{\alpha}^{(U)}$ is the gauge transformation of
$V_{\alpha}$ with respect to $U$. The prime on the functional delta
function means that the constant mode is excluded.  By the standard
argument, once this factor is inserted into the functional integral, a
gauge transformation allows one to factor out the volume factor
$\int(dU)$ which one can then divide out, and one is left with just
the Jacobian factor $\Delta$. $f$ in \fp\ is some arbitrary function
which can be integrated over with gaussian weight to induce the standard
gauge fixing term \gfix .

To evaluate $\Delta$ we expand the unitary integral about the point
where the argument of the delta function is zero. We then have, under
a gauge transformation,
\eqn\delV{
\delta V_{\alpha}=ig_sU^{\dagger}D_{\alpha}U
=D_{\alpha}\Lambda
\quad{\rm with}\quad
U=e^{-{i\over g_s}\Lambda}.}
We now write $\Lambda$, the unitary integral $\int(dU)$ and the
functional delta function in terms of modes. 
\eqn\mdexp{\eqalign{
&\Lambda(\sigma)=\sum_n\lambda_n\Lambda_n(\sigma)
\quad{\rm with}\quad
\lambda_n={<\Lambda_n|\Lambda>\over<\Lambda_n|\Lambda_n>}\cr
&\int (dU)=\int (d\Lambda)=\int\prod_n d\lambda_n|\Lambda_n|\cr
&\delta'(A-B)=\prod_{n\neq 0}\delta(a_n-b_n){1\over |\Lambda_n|}.}
}
We have used the flat space definition of the measure \dXmes\ for the
functional integral over $\Lambda$. The normalization factors 
$|\Lambda_n|$ are given by
\eqn\nmLn{
|\Lambda_n|=<\Lambda_n|\Lambda_n>^{1\over 2}.}
In the expression for the delta function of \mdexp\ $A$ and $B$ are
arbitrary functions which we have expanded in terms of the modes
$\Lambda_n$, i.e. $A(\sigma)=\sum_na_n\Lambda_n(\sigma)$ and similarly
for $B$. Note that there is nothing special about the decomposition
into the modes $\Lambda_n$, any (orthogonal) set of modes can be used.

Substituting in the expressions \mdexp\ into \fp\ we find
\eqn\fpt{
1=\Delta\,\,\int d\lambda_0|\Lambda_0|\,\,
\prod_{n\neq 0}{<\Lambda_n|\Lambda_n>\over<\Lambda_n|D^2|\Lambda_n>}.}
The integral over $\lambda_0$ is decoupled from the functional delta
function and hence is not localized. It corresponds to a global $U(1)$
rotation and from \delV\ we see that it must be integrated over the
finite interval $[0,2\pi g_s]$. The product(determinant) in \fpt\ can
be represented in the standard way by introducing a ghost action and
integrating over all ghost modes except the zero mode. We thus have
\eqn\delres{
\Delta={1\over 2\pi g_s|<\Lambda_0|\Lambda_0>|^{1\over 2}}
\int(d\bar{c}dc)'e^{\int\Tr[\bar{c}D^2c]},}
where the prime indicates that the integral excludes
the zero mode. The combined contribution from the ghost and
gauge zero modes is thus
\eqn\ghga{
(2\pi g_s)^2\biggl[{{\rm det}<V_i|V_j>\over
<\Lambda_0|\Lambda_0>}\biggr]^{1\over 2}.}
The factor of $g_s^2$ is the correct $g_s$ dependence
expected from string theory with each branch point contributing a
factor of $g_s$. 

As was originally argued in \bbng\ the zero mode structure of the
abelian gauge and ghost fields will always generate the correct 
$g_s$ weight for the topology of the string diagram.

\subsec{Integration over fermion zero modes}
In section 6 we constructed eight non-trivial fermion zero modes which
tended to constant but different values asymptotically far down each
of the four strings (\tho\thtw\thth\thfo ). In addition there are eight 
constant fermion zero modes taking the same value on each string,
$\theta^N$. We thus see that the fermion zero mode down the $i$th
string can be written as 
\eqn\thipN{
\pmatrix{
\theta_i\cr
\bar{\theta}_i\cr
\tilde{\theta}_i\cr
\bar{\tilde{\theta}}_i\cr
}
={1\over N_i}\pmatrix{
p_i^*\theta^p+N_i\theta^N\cr
p_i\bar{\theta}^p+N_i\bar{\theta}^N\cr
p_i\tilde{\theta}^p+N_i\tilde{\theta}^N\cr
p_i^*\bar{\tilde{\theta}}^p+N_i\bar{\tilde{\theta}}^N\cr
}
}
We now integrate over the fermion zero modes including a graviton
wavefunction for each of the four strings. Since the left and right moving
sectors of the fermionic integrals are identical we only need
focus on one of them. We have the following zero mode integral for the
left moving sector
\eqn\fzmint{
{\cal N}\int \,d^8\theta
\,\,\prod_{i=1}^4\Psi^i(\xi_i,\theta_i,\bar{\theta}_i)
\quad\quad{\rm with}\quad\quad
\eqalign{\int \,d^8\theta=&
\int \,d^2\theta^p \,d^2\bar{\theta}^p 
\,d^2\theta^N \,d^2\bar{\theta}^N\cr
{\cal N}=
&\bigl|{\rm det}<\theta^n|\theta^n>\bigr|^{-{1\over 2}}}.
}
${\cal N}$ is the normalization for the zero mode integral. 
It follows from the fermionic equivalent of \dXmescond . 
The wavefunctions are given in \Lwvfnc\ and the zero modes
$\theta_i$ and $\bar{\theta}_i$ are given in terms of
$\theta^p$,$\bar{\theta}^p$,$\theta^N$, and $\bar{\theta}^N$ through
equation \thipN . The fermionic integrals will pick out from the
product of wavefunctions precisely two $\theta^p$, two
$\bar{\theta}^p$, two $\theta^N$ and two $\bar{\theta}^N$.
Using the fact that the wavefunctions break into three subspaces
\ottffsse\ the fermionic integrals pick out of the product of
wavefunctions six different types of contribution. In other words,
under the fermionic integrals we have the identity
\eqn\effprod{
\eqalign{
\prod_{i=1}^4\Psi^i(\xi_i,\theta_i,\bar{\theta}_i)
=&\prod_{i=1}^4\Psi^i_{(12)}
+\prod_{i=1}^4\Psi^i_{(34)}
+\prod_{i=1}^4\Psi^i_{(5678)}\cr
&\hskip20pt+\sum_{i\neq j\neq k \neq l}\bigl(
\Psi^i_{(12)}\Psi^j_{(12)}\Psi^k_{(34)}\Psi^l_{(34)}
+\Psi^i_{(12)}\Psi^j_{(12)}\Psi^k_{(5678)}\Psi^l_{(5678)}\cr
&\hskip190pt+\Psi^i_{(34)}\Psi^j_{(34)}\Psi^k_{(5678)}\Psi^l_{(5678)}
\bigr)
}
}
Each of these six contributions has to be evaluated separately. The
calculations are lengthy but straightforward. The appendix lists the
identities necessary to derive the results and presents the first
calculation in detail. It also gives essential intermediate results to
aid verification of the five other calculations. Although the
calculations are not particularly enlightening in themselves, the 
reader is nevertheless urged to turn to the appendix to convince
him(her)self that it is highly non-trivial that the fermion zero mode
integrations produce the ten dimensionally Lorentz invariant results
enumerated below. In this section we will simply state the results,
referring the reader to the appendix for all details of the
calculations.

We start by considering the case in which all polarization vectors are
in the $X^5$,$X^6$,$X^7$,$X^8$ space. Performing the fermionic
integrations we find, after several pages of algebra, (see appendix)
that we can express the final result in terms of Lorentz invariant
quatities. Specifically we find
\eqn\fssesq{
\int d^8\theta\prod_{i=1}^4\Psi^i_{(5678)}=
-\bigl[ut(\xi_1.\xi_2)(\xi_3.\xi_4)+st(\xi_1.\xi_3)(\xi_2.\xi_4)
+su(\xi_1.\xi_4)(\xi_2.\xi_4)\bigr],
}
where $s$, $t$ and $u$ are the ten dimensional Lorentz invariants
given in \stu\stuexp . This is of the form expected for all four
polarization vectors orthogonal to the momenta. (see for example
\Sch\gsw )

The next case considered is that in which all four polarization
tensors lie in the
$X^3$,$X^4$ space. Integrating over the grassman variables one finds,
after several pages of algebra (see the appendix for essential details)
that the result can also be expressed in terms of Lorentz invariants. The
final result reads
\eqn\tfsq{\eqalign{
\int d^8\theta\prod_{i=1}^4\Psi^i_{(34)}=&\cr&\hskip-40pt
4\bigl[
s^2\tilde{\xi}_1\tilde{\xi}_2\tilde{\xi}_3^*\tilde{\xi}_4^*
+u^2\tilde{\xi}_1\tilde{\xi}_3\tilde{\xi}_2^*\tilde{\xi}_4^*
+t^2\tilde{\xi}_1\tilde{\xi}_4\tilde{\xi}_2^*\tilde{\xi}_3^*
+t^2\tilde{\xi}_2\tilde{\xi}_3\tilde{\xi}_1^*\tilde{\xi}_4^*
+u^2\tilde{\xi}_2\tilde{\xi}_4\tilde{\xi}_1^*\tilde{\xi}_3^*
+s^2\tilde{\xi}_3\tilde{\xi}_4\tilde{\xi}_1^*\tilde{\xi}_2^*
\bigr].
}
}
Using the identity \stuid\ and the definitions for $\xi$ and
$\bar{\xi}$ of \ottffsse\ it is straightforward to show that
this is again of the form \fssesq .

Focusing next on the case with two polarizations in the
$X^5$,$X^6$,$X^7$,$X^8$ space and two in the $X^3$,$X^4$ space we find
that there are six types of contribution depending on the way the
pairs of polarizations are divided amongst the four
strings. Performing the fermionic integrations we find after a long
calculation (see the appendix for essential results) that the results
can be expressed in the Lorentz invariant form
\eqn\tffsse{\eqalign{
\int d^8\theta
&\Psi^1_{(34)}\Psi^2_{(34)}\Psi^3_{(5678)}\Psi^4_{(5678)}
=-2ut(\tilde{\xi}_1\tilde{\xi}_2^*+\tilde{\xi}_2\tilde{\xi}_1^*)
(\xi_3.\xi_4)\cr
\int d^8\theta
&\Psi^1_{(34)}\Psi^3_{(34)}\Psi^2_{(5678)}\Psi^4_{(5678)}
=-2st(\tilde{\xi}_1\tilde{\xi}_3^*+\tilde{\xi}_3\tilde{\xi}_1^*)
(\xi_2.\xi_4)\cr
\int d^8\theta
&\Psi^1_{(34)}\Psi^4_{(34)}\Psi^2_{(5678)}\Psi^3_{(5678)}
=-2su(\tilde{\xi}_1\tilde{\xi}_4^*+\tilde{\xi}_4\tilde{\xi}_1^*)
(\xi_2.\xi_3)\cr
\int d^8\theta
&\Psi^2_{(34)}\Psi^3_{(34)}\Psi^1_{(5678)}\Psi^4_{(5678)}
=-2su(\tilde{\xi}_2\tilde{\xi}_3^*+\tilde{\xi}_3\tilde{\xi}_2^*)
(\xi_1.\xi_4)\cr
\int d^8\theta
&\Psi^2_{(34)}\Psi^4_{(34)}\Psi^1_{(5678)}\Psi^3_{(5678)}
=-2st(\tilde{\xi}_2\tilde{\xi}_4^*+\tilde{\xi}_4\tilde{\xi}_2^*)
(\xi_1.\xi_3)\cr
\int d^8\theta
&\Psi^3_{(34)}\Psi^4_{(34)}\Psi^1_{(5678)}\Psi^2_{(5678)}
=-2ut(\tilde{\xi}_3\tilde{\xi}_4^*+\tilde{\xi}_4\tilde{\xi}_3^*)
(\xi_1.\xi_2),\cr
}
}
where, $s$,$t$ and $u$ are the Lorentz invariants defined in
\stuexp . This is again of the form \fssesq .
In conclusion we see that for arbitrary polarization vectors lying in
the $X^3$,$X^4$,$X^5$,$X^6$,$X^7$,$X^8$ space, i.e. transverse to the
momenta, the result can always be written in the form \fssesq .

We still have to evaluate the fermion integrals for the three cases
where two or more polarizations are in the $X^1$,$X^2$ plane. We begin
with the case in which all four polarizations lie in the $X^1$,$X^2$
plane. Evaluating the fermion integrals we find a relatively simple
result (see the appendix). The result can then be shown to be equal to
the Lorentz invariant quantity given below.
\eqn\otot{\eqalign{
\int d^8\theta \prod_{i=1}^4\Psi^i_{(12)}
=&-\bigl[ut(\xi_1.\xi_2)(\xi_3.\xi_4)+st(\xi_1.\xi_3)(\xi_2.\xi_4)
+su(\xi_1.\xi_4)(\xi_2.\xi_3)\bigr]\cr
&\hskip10pt+2s\bigl[(\xi_1.k_4)(\xi_3.k_2)(\xi_2.\xi_4)
         +(\xi_2.k_3)(\xi_4.k_1)(\xi_1.\xi_3)\cr
&\hskip35pt+(\xi_1.k_3)(\xi_4.k_2)(\xi_2.\xi_3)
         +(\xi_2.k_4)(\xi_3.k_1)(\xi_1.\xi_4)\bigr]\cr
&\hskip10pt+2t\bigl[(\xi_2.k_1)(\xi_4.k_3)(\xi_1.\xi_3)
         +(\xi_3.k_4)(\xi_1.k_2)(\xi_2.\xi_4)\cr
&\hskip35pt+(\xi_2.k_4)(\xi_1.k_3)(\xi_3.\xi_4)
         +(\xi_3.k_1)(\xi_4.k_2)(\xi_1.\xi_2)\bigr]\cr
&\hskip10pt+2u\bigl[(\xi_1.k_2)(\xi_4.k_3)(\xi_2.\xi_3)
         +(\xi_3.k_4)(\xi_2.k_1)(\xi_1.\xi_4)\cr
&\hskip35pt+(\xi_1.k_4)(\xi_2.k_3)(\xi_3.\xi_4)
         +(\xi_3.k_4)(\xi_4.k_1)(\xi_1.\xi_2)\bigr]\cr}.}
The vector products are the Lorentz invariant vector products, with
$k_i$ the ten vectors of momenta \mom\ and $\xi$ the ten vectors
of polarization which include a $\xi^-$ part given by equation \xipxim
. To verify equation \otot\ one substitutes into the right hand side
the mometum ten vectors \mom\ and polarization ten vectors
\physpol\xipxim . The complicated form of the right hand side of
\otot\ can then be shown to reduce down to the simple result given in
the appendix. It is recommended to use a computer algebra program such
as Mathematica to check this.

The result \otot\ is precisely the four graviton kinematic factor
calculated in string theory \Sch\gsw . It is totally symmetric under
interchange of the strings and is furthermore gauge invariant. In
other words if one replaces any one polarization tensor by the
corresponding momenta the result is zero.

Next we treat the cases in which two polarizations lie in the
$X^1$,$X^2$ plane and two in the $X^3$,$X^4$ plane. There are six
different possible ways of distributing the two types of polarization
amongst the four strings. Computing the fermionic integrals one
arrives at some relatively simple expressions (see the appendix). It
is then straightforward to show that they can be expressed in the
Lorentz invariant form listed below.
\eqn\ottf{\eqalign{
&\int d^8\theta \Psi^1_{(12)}\Psi^2_{(12)}\Psi^3_{(34)}\Psi^4_{(34)}
=-\bigl[ut(\xi_1.\xi_2)-2t(\xi_1.k_3)(\xi_2.k_4)
  -2u(\xi_1.k_4)(\xi_2.k_3)\bigr](\xi_3.\xi_4)\cr
&\int d^8\theta \Psi^1_{(12)}\Psi^3_{(12)}\Psi^2_{(34)}\Psi^4_{(34)}
=-\bigl[st(\xi_1.\xi_3)-2s(\xi_1.k_4)(\xi_3.k_2)
  -2t(\xi_3.k_4)(\xi_1.k_2)\bigr](\xi_2.\xi_4)\cr
&\int d^8\theta \Psi^1_{(12)}\Psi^4_{(12)}\Psi^2_{(34)}\Psi^3_{(34)}
=-\bigl[us(\xi_1.\xi_4)-2s(\xi_1.k_3)(\xi_4.k_2)
  -2u(\xi_1.k_2)(\xi_4.k_3)\bigr](\xi_2.\xi_3)\cr
&\int d^8\theta \Psi^2_{(12)}\Psi^3_{(12)}\Psi^1_{(34)}\Psi^4_{(34)}
=-\bigl[us(\xi_2.\xi_3)-2s(\xi_2.k_4)(\xi_3.k_1)
  -2u(\xi_3.k_4)(\xi_2.k_1)\bigr](\xi_1.\xi_4)\cr
&\int d^8\theta \Psi^2_{(12)}\Psi^4_{(12)}\Psi^1_{(34)}\Psi^3_{(34)}
=-\bigl[st(\xi_2.\xi_4)-2s(\xi_2.k_3)(\xi_4.k_1)
  -2t(\xi_2.k_1)(\xi_4.k_3)\bigr](\xi_1.\xi_3)\cr
&\int d^8\theta \Psi^3_{(12)}\Psi^4_{(12)}\Psi^1_{(34)}\Psi^2_{(34)}
=-\bigl[ut(\xi_3.\xi_4)-2t(\xi_3.k_1)(\xi_4.k_2)
  -2u(\xi_3.k_2)(\xi_4.k_1)\bigr](\xi_1.\xi_2)\cr}.}
These are again of the form \otot .

Finally for the case with two polarizations in the $X^1$,$X^2$ plane 
and two in the subspace $X^5$,$X^6$,$X^7$,$X^8$ we find (see appendix)
identical results to those of \ottf\ with the wavefunctions
$\Psi_{(34)}$ replaced by $\Psi_{(5678)}$.

The result of this analysis is that an
arbitrary physical polarization vector, determined by its components in
the eight dimensional transverse space, leads to the Lorentz invariant
result of equation \otot . The restoration of 
the full $SO(8)$ invariance of the theory was to be expected from a
correct quantization of the theory, but the extension of this
invariance to ten dimensional Lorentz invariance seems to be a 
very non-trivial test of matrix string theory. 

\subsec{The four graviton scattering amplitude}
Combining the integrals over the left and right moving
fermion zero modes, the bosonic and gauge zero mode integrals
and the determinant from the non zero modes
of section 4 we arrive at the final result for the four string
scattering amplitude~:
\eqn\fgrsc{
{\cal A}={\cal I}\,\,{g_s^2\over stu}\,\, 
\,\,e^{-{1\over 4}(s\log s+t\log t+u\log u)}\,\,
{\cal K}_{\mu_1\mu_2\mu_3\mu_4}{\cal K}_{\nu_1\nu_2\nu_3\nu_4}
\epsilon_1^{\mu_1\nu_1}\epsilon_2^{\mu_2\nu_2}
\epsilon_3^{\mu_3\nu_3}\epsilon_4^{\mu_4\nu_4},}
where the tensor ${\cal K}_{\mu_1\mu_2\mu_3\mu_4}$ is the coefficient of 
$\xi_1^{\mu_1}\xi_2^{\mu_2}\xi_3^{\mu_3}\xi_4^{\mu_4}$ in equation
\otot .
The factor ${\cal I}$ contains the finite determinants from the zero
mode normalizations along with the fluctuation determinant ${\cal J}$
of section 4. Up to an overall numerical factor it is given by 
\eqn\calIdets{
{\cal I}=\,\,c\,\,{\cal J}\,\,\biggl[{
<\Lambda_0|\Lambda_0>{\rm det}<\theta^n|\theta^n>
\over
{\rm det}<V_i|V_j>{\rm det}<Y_q|Y_r>
}\biggr]^{1\over 2},}
where $c$ is the non-Lorentz invariant quantity defined in \defc .
We have shown in section 4 that the fluctuation determinant ${\cal J}$
does not receive any singular contributions from the branch points. We
further argued that up to small corrections the bosonic and fermionic
determinants would cancel, but we do not know how to evaluate 
${\cal J}$ more precisely. We can, however, take some inspiration from
conformal field theory.  In CFT it is well known that the finite
determinants for the zero modes combine with determinants from the
non-zero modes to give a result (in the critical dimension) that is
conformally invariant.  The heat kernel methods used to show this (see
for example \hp ) do not generalise in any obvious way to the
determinants calculated in this paper.  It seems likely, however, that
a similar mechanism is at work for the combination of determinants of
\calIdets .\footnote{$^5$}{I thank Herman Verlinde for emphasising
the importance of this effect in CFT and for suggesting that a similar
mechanism could play a role here.} As already discussed at the end of
section (8.1) it is also quite possible that the non-Lorentz invariant
factor $c$ would also be absorbed in a natural way into the
determinant (again this is precisely what happens in the string theory
calculation for the same amplitude).

If this is the case then, up to an overall numerical
factor, we have reproduced from matrix string theory the high energy
limit of the string theory
scattering amplitude for four gravitons \Sch\gsw\gmI\gmII .

\newsec{Discussion and conclusions}
In the previous section we have performed an analysis of graviton
scattering.  This formalism can however be extended in a
straightforward way to wavefunctions involving fermionic ground states
for the left and/or right moving sectors. These would be constructed
from odd numbers of $\theta$, and would permit the construction of the
complete multiplet of massless states of the type IIA superstring.

\subsec{Summary of the general calculational procedure}
We study scattering amplitudes which in string theory
are dominated by the points in moduli space corresponding to saddle
points of the classical action. We focused on the tree level
contribution, but in string theory there is a whole tower of saddle point
world sheets of higher topology that also
contribute.\footnote{$^6$}{In the original analysis of Gross
and Mende \gmI\gmII\ it was found that the perturbation series was
strongly divergent with genus $g$ world sheets giving a contribution
proportional to $g^{9g}$.} The methods developed in this
paper should be directly applicable to these higher genus
contributions. Let us summarize the general procedure.

The classical matrix string solutions corresponding to the saddle point
world sheets are used as background field
configurations. The quantum fluctuations around the 
backgrounds are treated by a one loop calculation. This consists of a
zero mode part and a fluctuation determinant for the non-zero
modes. All the essential structure of the amplitude is contained in
the zero mode integrations which can be calculated exactly. To find
the zero modes one first uses the fact that around the interaction
points there are instanton like field configurations. Locally these
break the translation symmetry and part of the supersymmetry. Each
broken symmetry has a corresponding zero
mode. They are given by equations \trzm\fsyzm 
\eqn\bfzmconcl{
V_{\mu}=F_{\alpha\mu}
\quad{\rm and}\quad
\theta=F_{\mu\nu}\Gamma_{\mu\nu}\epsilon.}
where $F_{\mu\nu}$ is the background field strength written in ten
dimensional notation (see section 3.2).
In addition there are
zero modes for the two dimensional gauge field. They can be written as 
\eqn\gfzmconcl{
V_{\mu}=D_{\mu}X_I,}
where $X_I$ is any of the non zero background bosonic fields. The zero
mode field configurations \bfzmconcl\gfzmconcl\ are finite at the
interaction points. Since, however, the background bosonic field
configurations $X_I$ correspond to world sheets branch points, the
zero modes \bfzmconcl\gfzmconcl\ will behave, asymptotically far from
the branch point, like singular string world sheet zero modes, with the
singularity sitting at the branch point. These modes can be glued into
global world sheet zero modes. 

The important point about the construction of these global zero modes
is that they only depend on the existence of a finite instanton
like classical solution around the branch point, not on the precise details
of the field configuration. 

For any world sheet it should be possible to identify,
using \bfzmconcl\gfzmconcl , the allowed singularities of the
zero modes at the branch points and hence to construct the global zero
modes.

Integration over the gauge field zero modes, along with the single,
constant, ghost field zero mode, leads to the correct power of $g_s$ for
the string world sheet\bbng .  The global bosonic zero modes
correspond to the collective coordinates for the positions of the
branch points. Integration over the global fermion zero modes gives
the ten dimensionally Lorentz invariant kinematic structure of the
scattering amplitude. This we have has explicitly verified for the
case of four graviton scattering.

\subsec{Connection with light cone superstring calculations}
The matrix string calculations described above have a strong
resemblance to the functional integral methods used in light cone
superstring calculations (see the review article \rtI ). Below we
discuss the similarities and differences between the two. 

In light cone superstring calculations the functional integral reduces
down to an integration over zero modes. There are, however, no zero
modes corresponding to the translation of the branch point and no
abelian gauge field and so no gauge or ghost field zero modes. The
inclusion of factors of $g_s$ and the integrals over the string moduli
are added by hand.

To perform the Euclidean functional integral the Minkowski space
Majorana-Weyl fermions are combined into complex fermionic
coordinates and their conjugate momenta. These can be defined by the
same formula as for matrix string theory i.e. equation \defth .  This
breaks the $spin(8)$ symmetry down to $U(1)\otimes SU(4)$. The
$\theta$ and $\bar{\theta}$ of \defth\ form a ${\bf 4}$ of $SU(4)$
and the $\tilde{\theta}$ and $\bar{\tilde{\theta}}$ form a ${\bf \bar{4}}$.

In the matrix string theory calculation, there is a classical background,
that through the non-abelian nature of the theory, interacts with the
fermions and further breaks the $SU(4)$ symmetry. In the four 
string scattering process discussed in this paper the original
$spin(8)$ symmetry is broken down to $U(1)\otimes U(1)\otimes
SU(2)\otimes SU(2)$. Other classical backgrounds, for example
five or six string scattering, would break the $spin(8)$ symmetry even
further.

In light cone superstring calculations the fermionic coordinates
$\theta$ and momenta $\lambda$ have the conformal weights $1$ and $0$
respectively. The functional integral over the fermions leads to a
determinant and 
to an integration over the $\theta$ zero modes. These are given by the
complete set of abelian differentials defined on the string diagram
Riemann surface. In terms of the light cone world
sheet coordinate $w$, they consist of all the modes satisfying
\eqn\lcfzm{
\partial_w\theta=0.}
We denote by $\theta$ in \lcfzm\ the $\bf 4$ of $SU(4)$ made from
$\theta$ and $\bar{\theta}$ of \defth . There is also the complex
conjugate equation for the $\bf \bar{4}$.  The modes satisfying
\lcfzm\ are allowed to have $1/\sqrt{w-w_0}$ singularities at each
simple branch point $w_0$, and have the boundary condition that they
tend to constant values asymptotically far down the strings.  The
generic fermion zero mode thus has a singularity at each branch point.
This is in contrast to the matrix string zero modes of section 6.3
which each had a single singularity even though there were two branch
points. In other words, for this process, there are twice as many
fermion zero modes in the light cone superstring calculation as there
are in the matrix string calculation.

This mismatch is resolved by the final element of light cone
superstring calculations~: the addition by hand, at each branch point
of an operator ${\cal V}$ needed to preserve supersymmetry. The
precise form of this operator for simple branch points (i.e. where two
strings interact) was calculated in
\gs . It can be written in the form
\eqn\lcop{
{\cal V}=-{4\over|2\partial^2_zw(z)|}\,\,
\Psi(\partial_zX^I,\partial_z\theta,\partial_z\bar{\theta})\,\,
\tilde{\Psi}(\partial_{\bar{z}}X^I,\partial_{\bar{z}}\tilde{\theta},
\partial_{\bar{z}}\bar{\tilde{\theta}}).}
where $\Psi(\,\,,\,\,,\,\,)$ is as in \Lwvfnc\ except with $p+$
replaced by $1/(2\partial^2_zw(z))$, and in the right moving part
$\tilde{\Psi}(\,\,,\,\,,\,\,)$ $p^+$ is replaced by
$1/(2\partial^2_zw(z))^*$.  $w$ is the light cone coordinate expressed,
through the Mandelstam mapping, in terms of the uniformization $z$ (for
tree level scattering see equation \sphwz ). The bosonic fields $X$ in
the operators ${\cal V}$ are contracted both with the external
wavefunctions and with the other ${\cal V}$. The fermions $\theta$ are
the fermion zero modes.  It is straightforward to see that the two
interaction operators used for the four string scattering amplitude
soak up the excess zero modes.

\subsec{Conclusions and future directions}
The matrix string zero mode calculation of this paper is considerably
simpler than that for the light cone superstring, where an interaction
operator has to be introduced by hand at each interaction point.
So far, however, the matrix string calculations have only been carried out
about a classical background corresponding to a saddle point
of the classical action. It is important to extend them 
to other backgrounds. In the context of the four-string scattering
calculation, it is clear that there is a problem when the world
sheet is no longer a stationary point of the action. In this case, the
world sheet is not (anti)holomorphic around the branch
points and the corresponding instanton solution will break
all of the supersymmetry leading to extra fermion zero modes. There
must be a simple mechanism that soaks up these extra zero modes. It
is important to understand this as it could permit the
calculations to be done at an arbitrary point in moduli space.

The semi-classical analysis of this paper has been justified by
focusing on very high energy scattering processes. Away from this
limit we do not know how to calculate. There will however be an
effective action describing the low energy fluctuations, even if we do
not at present have the technical tools to obtain it. The zero
mode analysis of this paper only depends on the existence of a finite
field configuration for the interaction points, and should also be
applicable to this effective field theory.

Finally it is interesting to speculate on how the semi-classical
reasoning of this paper could be made more rigorous. In simpler
situations, semiclassical methods have recently been shown to be in
precise agreement with exact results. Using quasi-classics,
the partition function for matrix string theory on a torus was
calculated \kvh\ and shown to agree, on extrapolation to the small torus
limit, with the exact results for the completely reduced SYM theory
\mns . These latter results were obtained by mapping the original
SYM theory to a cohomological field theory (CohFT) using the methods of
\witt . Recently CohFT methods have been applied directly to the
matrix string partition function \fs\ beautifully confirming the
semi-classical results of \kvh . A key aspect of CohFT is that the
functional integral localizes on configurations of the classical
vacua. Although it is not possible to apply the twisting procedures
used in \witt\mns\fs\ to matrix string scattering processes it is
tempting to speculate that some kind of localization mechanism is also
behind the success of the calculations presented in this paper.

\vskip 30pt
\hskip -19pt{\bf Acknowledgements}
I would like to thank Philippe Brax, Frank Ferrari, Ivan Kostov,
and especially Herman Verlinde for many helpful discussions. 
I also thank the IHES for their hospitality where part of this
work was carried out.

\appendix{1}{Gamma matrix definitions}
We choose as the basis of real spin(8) gamma matrices $\Gamma^I$ and
Majorana-Weyl fermions $S$
\eqn\gamgam{
\Gamma^0=\pmatrix{1&0\cr 0&1},
\Gamma^I=\pmatrix{0&\gamma^I_{a\dot{a}}\cr\gamma^I_{\dot{a}a}&0\cr},
\Gamma^9=\pmatrix{1&0\cr 0&-1},
\quad\quad
S=\pmatrix{S^a\cr S^{\dot{a}}},
}
where the indices $a$ and $\dot{a}$ which run from one to eight label
the {\bf 8}$_s$ and {\bf 8}$_c$ representations of $spin(8)$ respectively.
The $\gamma^I_{a\dot{a}}$ (which are the transposes of 
the $\gamma^I_{\dot{a}a}$) are given by
\eqn\gammadef{
\eqalign{
\gamma^1=&\,\,1\otimes1\otimes1\cr
\gamma^3=&\,\,\tau_1\otimes\tau_3\otimes\epsilon\cr
\gamma^5=&\,\,\tau_1\otimes\epsilon\otimes1\cr
\gamma^7=&-\tau_3\otimes1\otimes\epsilon
}
\quad\quad
\eqalign{
\gamma^2=&\,\,\epsilon\otimes1\otimes1\cr
\gamma^4=&\,\,\tau_1\otimes\tau_1\otimes\epsilon\cr
\gamma^6=&\,\,\tau_3\otimes\epsilon\otimes\tau_1\cr
\gamma^8=&\,\,\tau_3\otimes\epsilon\otimes\tau_3,
}
}
where $\epsilon=i\tau_2$ and $\tau_1$,$\tau_2$ and $\tau_3$ are the
pauli matrices.
With this basis the gamma matrices $\Gamma^0$,$\Gamma^9$,$\Gamma^1$ and
$\Gamma^2$ are given by
\eqn\gonot{
\Gamma^0=\pmatrix{1&0\cr 0&1\cr}\quad
\Gamma^9=\pmatrix{1&0\cr 0&-1\cr}\quad
\Gamma^1=\pmatrix{0&1\cr 1&0\cr}\quad
\Gamma^2=\pmatrix{0&i\tau_2\cr -i\tau_2&0\cr},
}
where all entries correspond to 8x8 blocks with
the pauli matrix $\tau_2$ being written as 4x4 blocks.

With respect to these four gamma matrices the sixteen component
Majorana-Weyl spinors thus decompose into four blocks of four. We
will be working in Euclidean space where ten dimensional Majorana-Weyl
spinors do not exist. Furthermore, as discussed in section 3 
each Majorana-Weyl fermionic variable is simultaneously a fermionic
coordinate and its conjugate momenta. We thus have to combine the fermionic
variables into complex fermionic coordinates and their
distinct conjugate momenta. It is thus 
convenient to work with a complex basis of gamma matrices defined by 
\eqn\compgam{
\gamma\rightarrow 
(v\otimes v\otimes1)\gamma
(v^{\dagger}\otimes v^{\dagger}\otimes1)
\quad{\rm with}\quad
v={1\over\sqrt{2}}\pmatrix{1&i\cr 1&-i}.
}
Written out explicitly they are given by
\eqn\compgammadef{
\eqalign{
\gamma^1=&\,\,1\otimes1\otimes1\cr
\gamma^3=&-i\tau_2\otimes\tau_1\otimes\tau_2\cr
\gamma^5=&\,\,i\tau_2\otimes\tau_3\otimes1\cr
\gamma^7=&-\tau_1\otimes1\otimes\tau_2
}
\quad\quad
\eqalign{
\gamma^2=&-i\tau_3\otimes1\otimes1\cr
\gamma^4=&\,\,i\tau_2\otimes\tau_2\otimes\tau_2\cr
\gamma^6=&-i\tau_1\otimes\tau_3\otimes\tau_1\cr
\gamma^8=&-i\tau_1\otimes\tau_3\otimes\tau_3,
}
}
In the basis of \compgammadef\ the four gamma matrices that appear in
the background covariant Dirac operator are identical to those
of \gonot\ except for $\Gamma^2$ where $\tau_2$ is replaced by
$-\tau_3$. Finally it is useful for the study of the fermion zero modes and 
determinant to put the Dirac operator into 
block diagonal form  by mixing the two $spin(8)$ 
representations. Specifically, writing the sixteen by sixteen $\Gamma$
matrices as four by four blocks, we interchange the second and third
rows and columns. This leads to the following block diagonal form for
the gamma matrices $\Gamma^0$,$\Gamma^9$,$\Gamma^1$ and  
$\Gamma^2$ 
\eqn\gonot{
\Gamma^0=\pmatrix{1&0\cr 0&1\cr}\quad
\Gamma^9=\pmatrix{\sigma_3&0\cr 0&\sigma_3\cr}\quad
\Gamma^1=\pmatrix{-\sigma_2&0\cr 0&\sigma_2\cr}\quad
\Gamma^2=\pmatrix{\sigma_1&0\cr 0&\sigma_1\cr},
}
where again all entries are 8x8 blocks.  This transformation also
interchanges the $\bar{\theta}_{\bar{A}}$,$\lambda_A$ with
$\tilde{\theta}^A$,$\bar{\tilde{\lambda}}^{\bar{A}}$ in (2.1) of
appendix 2 and
leads to the block diagonal form for the Euclidean fermion action
given in \fermL .
\appendix{2}{Fermionic coordinates and momenta}
Under the change of basis \compgam\ the Majorana-Weyl fermions split into 
fermionic coordinates $\theta$ and momenta $\lambda$ as follows
\eqn\Sgotothl{
S^a\rightarrow(v\otimes v\otimes1)S^a=
\pmatrix{\theta^A\cr
\bar{\lambda}^{\bar{A}}\cr
\bar{\theta}_{\bar{A}}\cr
\lambda_A\cr}
\quad{\rm and}\quad
S^{\dot{a}}\rightarrow(v\otimes v\otimes1)S^{\dot{a}}=
\pmatrix{\tilde{\theta}^A\cr
\bar{\tilde{\lambda}}^{\bar{A}}\cr
\bar{\tilde{\theta}}_{\bar{A}}\cr
\tilde{\lambda}_A\cr},
}
where the indices $A$,$\bar{A}$ take the values $1,2$
with the only non-zero anticommutators being
\eqn\comrel{
\eqalign{
\{\theta^A,\lambda_B\}=&\delta^A_B\cr
\{\bar{\theta}_{\bar{A}},\bar{\lambda}^{\bar{B}}\}
=&\delta_{\bar{A}}^{\bar{B}}
}
\quad\quad{\rm and}\quad\quad
\eqalign{
\{\tilde{\theta}^A,\tilde{\lambda}_B\}=&\delta^A_B\cr
\{\bar{\tilde{\theta}}_{\bar{A}},\bar{\tilde{\lambda}}^{\bar{B}}\}
=&\delta_{\bar{A}}^{\bar{B}},
}
}
where we have not explicitly included the spatial delta functions.
The transformation of the fermion coordinates and momenta under
rotations in the $X^1$,$X^2$ and $X^3$,$X^4$ planes and in the four
dimensional space $X^5$,$X^6$,$X^7$ and $X^8$ can be read off from
the explicit form for the rotation generators in the basis
\compgammadef . Specifically under an infinitesimal rotation
described by $w^{IJ}$ we have 
\eqn\delS{
\eqalign{
\delta S^a=&{1\over 4}w^{IJ}\gamma^{IJ}_{ab} S^b\cr
\delta S^{\dot{a}}=
&{1\over 4}w^{IJ}\tilde{\gamma}^{IJ}_{\dot{a}\dot{b}}S^{\dot{b}}
}
\quad\quad{\rm with}\quad\quad
\eqalign{
\gamma^{IJ}_{ab}=&
{1\over
2}(\gamma^I(\gamma^J)^{\dagger}-\gamma^J(\gamma^I)^{\dagger})\cr
\tilde{\gamma}^{IJ}_{\dot{a}\dot{b}}=&
{1\over 2}((\gamma^I)^{\dagger}\gamma^J-(\gamma^J)^{\dagger}\gamma^I)
}
}
We can thus draw up the following table for the action of 
$\gamma^{IJ}$ on $\theta^A$,$\bar{\theta}_{\bar{A}}$,$\lambda_A$ 
and $\bar{\lambda}^{\bar{A}}$
\eqn\transtb{
\vbox{\offinterlineskip
\halign{\quad#\quad&\vrule\quad#\quad&\quad#\quad&\quad#\quad&\quad#\quad\cr
\omit&\omit&\omit&\omit&\omit\cr
\omit&$\theta^A$&$\tilde{\theta}_{\bar{A}}$&
$\lambda_A$&$\bar{\lambda}^{\bar{A}}$\cr
\omit&\omit&\omit&\omit&\omit\cr
\noalign{\hrule}
\omit&\omit&\omit&\omit&\omit\cr
$\gamma^{12}=\,\,\tau_3\otimes1\otimes i$&$i$&$-i$&$-i$&$i$\cr
$\gamma^{34}=\,\,1\otimes\tau_3\otimes i$&$i$&$i$&$-i$&$-i$\cr
$\gamma^{56}=\,\,\tau_3\otimes1\otimes i\tau_1$
&$i\tau_1$&$-i\tau_1$&$-i\tau_1$&$i\tau_1$\cr
$\gamma^{57}=\,\,\tau_3\otimes\tau_3\otimes i\tau_2$
&$i\tau_2$&$-i\tau_2$&$i\tau_2$&$-i\tau_2$\cr
$\gamma^{58}=\,\,\tau_3\otimes1\otimes i\tau_3$&
$i\tau_3$&$-i\tau_3$&$-i\tau_3$&$i\tau_3$\cr
$\gamma^{67}=\,\,1\otimes\tau_3\otimes i\tau_3$&
$i\tau_3$&$i\tau_3$&$-i\tau_3$&$-i\tau_3$\cr
$\gamma^{68}=\,\,1\otimes1\otimes -i\tau_2$&
$-i\tau_2$&$-i\tau_2$&$-i\tau_2$&$-i\tau_2$\cr
$\gamma^{78}=\,\,1\otimes\tau_3\otimes i\tau_1$&
$i\tau_1$&$i\tau_1$&$-i\tau_1$&$-i\tau_1$\cr}}
}
There is an almost identical table for the action of
$\tilde{\gamma}^{IJ}$ on $\tilde{\theta}^A$,
$\bar{\tilde{\theta}}_{\bar{A}}$,$\tilde{\lambda}_A$ 
and $\bar{\tilde{\lambda}}^{\bar{A}}$. The only difference from 
the above table being that all entries of the first line change sign.
We can read of directly from \transtb\ the $U(1)$ charges for 
rotations in the $X^1$,$X^2$ and $X^3$,$X^4$ planes. 
Further more we can read off the effect of infinitesimal $SO(4)$ 
rotations in the $X^5$,$X^6$,$X^7$ and $X^8$ subspace.
Using the notations of \West\ for four four dimensional spinors we have
\eqn\ththlltr{
\eqalign{
\delta\theta^A={1\over 4}w^{mn}(\sigma^{mn})^A_{\,\,\,\,B}\theta^B\cr
\delta\bar{\theta}_{\bar{A}}=
{1\over 4}w^{mn}(\bar{\sigma}^{mn})_{\bar{A}}^{\,\,\,\,\bar{B}}
\bar{\theta}_{\bar{B}}
}
\quad\quad
\eqalign{
\delta\lambda_A=-{1\over 4}w^{mn}(\sigma^{mn})^B_{\,\,\,\,A}\theta_B\cr
\delta\bar{\lambda}^{\bar{A}}=
-{1\over 4}w^{mn}(\bar{\sigma}^{mn})_{\bar{B}}^{\,\,\,\,\bar{A}}
\bar{\lambda}^{\bar{B}}.
}
}
where $(\sigma^{mn})^A_{\,\,\,\,B}$ and 
$(\bar{\sigma}^{mn})_{\bar{A}}^{\,\,\,\,\bar{B}}$ are defined by
\eqn\defsmn{
(\sigma^{mn})^A_{\,\,\,\,B}=
{1\over 2}(\sigma^m\bar{\sigma}^n-\sigma^n\bar{\sigma}^m)
\quad{\rm and}\quad
(\bar{\sigma}^{mn})_{\bar{A}}^{\,\,\,\,\bar{B}}=
{1\over 2}(\bar{\sigma}^m\sigma^n-\bar{\sigma}^n\sigma^m).
}
with $\sigma^m$ and $\bar{\sigma}^m$ defined through the Pauli matrices
\eqn\defsst{
(\sigma^m)^{A\bar{B}}=(i,\tau_1,\tau_2,\tau_3)
\quad{\rm and}\quad
(\bar{\sigma}^m)_{\bar{A}B}=(-i,\tau_1,\tau_2,\tau_3).}
\subsec{Identities for two component 4$d$ spinors}
We use the conventions of \West , modified by a factor of $i$ for the
zeroth component since the space $X^5$,$X^6$,$X^7$,$X^8$ has Euclidean 
metric. 

There are two pairs of two component spinors, $\theta^A$ and
$\bar{\theta}_{\bar{B}}$ with $A,\bar{B}=1,2$. Indices are raised and
lowered using the antisymmetric tensor $\epsilon$~:
\eqn\raislowind{
\theta^A=\epsilon^{AB}\theta_B,
\quad
\theta_A=\epsilon_{BA}\theta^B,
\quad
\bar{\theta}^{\bar{A}}=\epsilon^{\bar{A}\bar{B}}\bar{\theta}_{\bar{B}},
\quad
\bar{\theta}_{\bar{A}}=\epsilon^{\bar{B}\bar{A}}\bar{\theta}^{\bar{B}},
}
where the $\epsilon$ tensor is definied to be
\eqn\defep{
\epsilon_{AB}=\epsilon^{AB}
=-\epsilon_{\bar{A}\bar{B}}=-\epsilon^{\bar{A}\bar{B}}
\quad{\rm with}\quad
\epsilon_{AB}=\pmatrix{0&1\cr -1&0}.
}
We define $\theta^2$ and $\bar{\theta}^2$ by 
\eqn\defthsq{
\theta^2=\theta^A\theta_A
\quad\quad{\rm and}\quad\quad
\bar{\theta}^2=\bar{\theta}^{\bar{A}}\bar{\theta}_{\bar{A}}.
}
Integrals over $\theta$ and $\bar{\theta}$ are defined by
\eqn\intth{
\int d^2\theta=\int d\theta_1d\theta_2
\quad{\rm and}\quad
\int d^2\bar{\theta}=\int d\bar{\theta}_2d\bar{\theta}_1.
}
We define the four matrices $(\sigma^m)^{C\bar{D}}$ with
$m=5,6,7,8$ and their barred partners by
\eqn\sigbsig{
\sigma^m=(i,\tau),
\quad\quad{\rm and}\quad\quad
\bar{\sigma}^m=(-i,\tau).
}
Finally we contract the $\sigma^m$ with $\theta$ and $\bar{\theta}$
to produce a four vector in the space $X^5$,$X^6$,$X^7$ and $X^8$~:
\eqn\fvec{
\theta\sigma^m\bar{\theta}=
\theta_A(\sigma^m)^{A\bar{B}}\bar{\theta}_{\bar{B}}.
}

Using the above definitions we can derive identities for
calculating the fermionic integrals of section 8. Below we list 
the complete set of identities needed.
\eqn\epepdel{
\epsilon^{AB}\epsilon_{CB}=\delta^A_C
\quad\quad{\rm and}\quad\quad
\epsilon^{\bar{A}\bar{B}}\epsilon_{\bar{C}\bar{B}}=\delta^{\bar{A}}_{\bar{C}},
}
\eqn\thsq{
\theta^2=-\epsilon^{AB}\theta_A\theta_B
\quad\quad{\rm and}\quad\quad
\bar{\theta}^2=
-\epsilon^{\bar{A}\bar{B}}\bar{\theta}_{\bar{A}}\bar{\theta}_{\bar{B}},
}
\eqn\thint{
\int d^2\theta\pmatrix{\theta^2\cr\theta_A\theta_B}=
\pmatrix{2\cr -\epsilon_{AB}}
\quad\quad{\rm and}\quad\quad
\int d^2\bar{\theta}\pmatrix{\bar{\theta}^2\cr
\bar{\theta}_{\bar{A}}\bar{\theta}_{\bar{B}}}=
\pmatrix{2\cr -\epsilon_{\bar{A}\bar{B}}},
}
\eqn\bsigm{
(\bar{\sigma}^m)_{\bar{B}A}=
(\sigma^m)^{C\bar{D}}\epsilon_{CA}\epsilon^{\bar{D}\bar{B}},
}
\eqn\sigsig{
(\sigma^m)^{A\bar{B}}(\bar{\sigma}^n)_{\bar{B}C}
+(\sigma^n)^{A\bar{B}}(\bar{\sigma}^m)_{\bar{B}C}
=2\delta^{nm}\delta^A_C,}
and
\eqn\trsig{
\Tr[\sigma^m\bar{\sigma}^n]=2\delta^{nm}.
}

\appendix{3}{Fermion zero mode integrals}
Using the identities of the previous section the calculations of the 
fermion integrals of section 8, are straightforward
but somewhat tedious. In this section we present one calculation
in detail. The other calculations can be evaluated in a similar
manner and so for them we just give some essential
intermediate results. 
\vskip 17.5pt
For all four polarizations lying in the $X^5$,$X^6$,$X^7$,$X^8$ space
we have
\eqn\fssesqa{
\int\,d^8\theta\,\prod_{1=1}^4\Psi^i_{(5678)}
={\xi_1^{m_1}\xi_2^{m_2}\xi_3^{m_3}\xi_4^{m_4}\over N_1N_2N_3N_4}
{\cal I}^{m_1m_2m_3m_4},}
where
\eqn\calI{
{\cal I}^{m_1m_2m_3m_4}=
\int\,d^2\theta^p d^2\bar{\theta}^p d^2\theta^N d^2\bar{\theta}^N
\,\prod_{i=1}^4
(p_i^*\theta^p+N_i\theta^N)(\sigma^{m_i})
(p_i\bar{\theta}^p+N_i\bar{\theta}^N).}
The integrals over the $\theta$ only gives a non-zero result for the terms 
from the product in which there are two
$\theta^p$, two $\bar{\theta}^p$, two $\theta^N$ and two
$\bar{\theta}^N$. In other words each non-zero contribution will
consist of two factors of transverse momenta $p$, two factors of
transverse momenta $p^*$ and four factors of $p^+$ momenta $N$.
We thus see that there are three different types of integrals to be
calculated according to how the four $p$'s and four $N$'s are
distributed amongst the four wavefunctions. Specifically we have
\eqn\calIIII{
{\cal I}={\cal I}_1+{\cal I}_2+{\cal I}_3}
where
${\cal I}_1$ consists of 6 (=4!/(2!2!)) contributions of the form 
$|p_i|^2|p_j|^2N_k^2N_l^2$, 
${\cal I}_2$ consists of 24 (=4!) contributions of the form 
$|p_i|^2p_j^*p_kN_jN_kN_l^2$ and  
${\cal I}_3$ consists of 6 (=4!/(2!2!)) contributions of the form 
$p_i^*p_j^*p_kp_lN_iN_jN_kN_l$. 
For these three different types of term the $\theta$ integrals lead to
the following contributions
\eqn\thcontro{
\eqalign{
|p_i|^2|p_j|^2N_k^2N_l^2\,\,\int\,d^8\theta\,\,\,
\theta^p\sigma^{m_i}\bar{\theta}^p\theta^p\sigma^{m_j}\bar{\theta}^p
\theta^N\sigma^{m_k}\bar{\theta}^N\theta^N\sigma^{m_l}\bar{\theta}^N&\cr
&\hskip-100pt=4|p_i|^2|p_j|^2N_k^2N_l^2\delta^{m_im_j}\delta^{m_km_l}\cr
|p_i|^2p_j^*p_kN_jN_kN_l^2\,\,\int\,d^8\theta\,\,\,
\theta^p\sigma^{m_i}\bar{\theta}^p\theta^p\sigma^{m_j}\bar{\theta}^N
\theta^N\sigma^{m_k}\bar{\theta}^p\theta^N\sigma^{m_l}\bar{\theta}^N\cr
&\hskip-200pt=-2|p_i|^2p_j^*p_kN_jN_kN_l^2
(\delta^{m_im_j}\delta^{m_km_l}
+\delta^{m_im_k}\delta^{m_jm_l}
-\delta^{m_im_l}\delta^{m_jm_k})\cr
p_i^*p_j^*p_kp_lN_iN_jN_kN_l\,\,\int\,d^8\theta\,\,\,
\theta^p\sigma^{m_i}\bar{\theta}^N\theta^p\sigma^{m_j}\bar{\theta}^N
\theta^N\sigma^{m_k}\bar{\theta}^p\theta^N\sigma^{m_l}\bar{\theta}^p\cr
&\hskip-100pt=4p_i^*p_j^*p_kp_lN_iN_jN_kN_l\delta^{m_im_j}\delta^{m_km_l}.\cr
}
}
Adding together all the different contributions of a particular type
and using the identities \consm ,\poptp\ and \stuexp\ we find
\eqn\calIo{
{\cal I}_1=4{p^4\over N^4}
N_1N_2N_3N_4\bigl[2N_1N_2N_3N_4{\cal A}
+(N_2^2N_4^2+N_1^2N_3^2){\cal B}
+(N_2^2N_3^2+N_1^2N_4^2){\cal C}\bigr],}
\eqn\calIt{\eqalign{
{\cal I}_2=4{p^2\over N^2}
N_1N_2N_3N_4\bigl[
&{p^2\over N^2}\bigl((N_1^2+N_2^2)N_3N_4+(N_3^2+N_4^2)N_1N_2\bigr)
(-{\cal A}+{\cal B}+{\cal C})\cr
&\hskip100pt
-q^2(N_1N_4+N_2N_3)({\cal A}-{\cal B}+{\cal C})\cr
&\hskip100pt
-q^2(N_1N_3+N_2N_4)({\cal A}+{\cal B}-{\cal C})
\bigr],\cr}
}
and
\eqn\calIth{
{\cal I}_3=4N_1N_2N_3N_4\bigl[q^4{\cal A}+2{p^4\over N^4}N_1N_2N_3N_4
(-{\cal A}+{\cal B}+{\cal C})\bigr].}
where we have defined the tensors ${\cal A}$,${\cal B}$ and ${\cal C}$
by 
\eqn\calabc{
{\cal A}=\delta^{m_1m_2}\delta^{m_3m_4},
\quad
{\cal B}=\delta^{m_1m_3}\delta^{m_2m_4},
\quad{\rm and}\quad
{\cal C}=\delta^{m_1m_4}\delta^{m_2m_3}.}
Adding these three results together to obtain ${\cal I}$ \calIIII ,
substituting into \fssesqa\ and using the definitions \stuexp\ we
obtain the result \fssesq .
\vskip 17.5pt
For all four polarizations lying in the $X^3$,$X^4$ space we have
\eqn\tftfa{
\int\,d^8\theta\,\prod_{i=1}^4\Psi^i_{(34)}
={\tilde{\xi}_1\tilde{\xi}_2\tilde{\xi}^*_3\tilde{\xi}^*_4\over
N_3^2N_4^2}
{\cal I}(1,2;3,4)
\quad+\,5{\rm \,\,other\,\,permutations},}
where 
\eqn\calJ{
{\cal I}(i,j;k,l)=\int\,d^8\theta\,\prod_{m=k,l}
(p_m^*\theta^p+N_m\theta^N)^2
(p_m\bar{\theta}^p+N_m\bar{\theta}^N)^2.}
Evaluating the fermionic integrals (there are five different types of
contribution) we find
\eqn\calJexp{\eqalign{
{\cal I}(i,j;k,l)=&\,\,16\bigl[|p_k|^4N_l^4
-2|p_k|^2(p_kp_l^*+p_k^*p_l)N_kN_l^3
+p_k^2(p_l^*)^2N_k^2N_l^2\bigr]+(k\leftrightarrow l)\cr
&+64|p_k|^2|p_l|^2N_k^2N_l^2.}}
Using the identities \consm ,\poptp\ and \stuexp\ to evaluate the
above result for the six possible permutations of \tftfa\ we obtain
the result \tfsq .
\vskip 17.5pt
For two polarizations lying in the $X^3$,$X^4$ plane and two lying in
the $X^5$,$X^6$,$X^7$,$X^8$ space we have 
\eqn\tffssea{
\int\,d^8\theta\,\Psi^i_{(34)}\Psi^j_{(34)}\Psi^k_{(5678)}\Psi^l_{(5678)}
={\tilde{\xi}_i\tilde{\xi}_j^*(\xi_k.\xi_l)\over N_i^2N_kN_l}
{\cal I}(i;j|k,l)
\quad+\quad {\rm c.c.}}
where
\eqn\calIt{\eqalign{
{\cal I}(i;j|k,l)\delta^{m_km_l}=&\cr
&\hskip-70pt\epsilon\,\,\int\,d^8\theta\,\,\,
(p_j^*\theta^p+N_j\theta^N)^2
(p_j\bar{\theta}^p+N_j\bar{\theta}^N)^2
\prod_{n=k,l}(p_n^*\theta^p+N_n\theta^N)(\sigma^{m_n})
(p_n\bar{\theta}^p+N_n\bar{\theta}^N),}}
where the sign factor $\epsilon$ is $+1$ if $k,l=1,2$ or $3,4$ and
$-1$ otherwise. This factor comes from the sign of the 
$p^+$ momenta in the definitions of the wavefunctions 
\ottffsse .
Integrating over the $\theta$ (there are ten different types of
contribution) leads to the result
\eqn\calItr{
\eqalign{
{\cal I}(i;j|k,l)=&-8\epsilon\biggl[|p_j|^4N_k^2N_l^2+N_j^4|p_k|^2|p_l|^2
+|p_j|^2|p_k|^2N_j^2N_l^2+|p_j|^2|p_l|^2N_j^2N_k^2\cr
&\hskip24pt+|p_j|^2(p_kp_l^*+p_k^*p_l)N_j^2N_kN_l
+(p_j^2p_k^*p_l^*+(p_j^*)^2p_kp_l)N_j^2N_kN_l\cr
&\hskip36pt-|p_j|^2(p_jp_k^*+p_j^*p_k)N_jN_kN_l^2
-|p_j|^2(p_jp_l^*+p_j^*p_l)N_jN_k^2N_l\cr
&\hskip60pt-|p_k|^2(p_jp_l^*+p_j^*p_l)N_j^3N_l
-|p_l|^2(p_jp_k^*+p_j^*p_k)N_j^3N_k\biggr]}}
Using the identities \consm ,\poptp\ and \stuexp\ to evaluate the
above result for the six possible combinations of pairs of
polarizations leads to the results listed in \tffsse .
\vskip 17.5pt
For all four polarizations lying in the $X^1$,$X^2$ plane we have to
evaluate the integral
\eqn\otota{
\int\,d^8\theta\,\prod_{i=1}^4\Psi^i_{(12)}
={\xi_1\xi_2\xi_3^*\xi_4^*\over N_1N_2N_3N_4}{\cal I}(1,2;3,4)
\quad+\,\,5{\rm\,\,other\,\,permutations},}
where
\eqn\calIotot{
{\cal I}(i,j;k,l)=
\int\,d^8\theta\,\prod_{m=i,j}\prod_{n=k,l}
(p_n^*\theta^p+N_n\theta^N)^2
(p_m\bar{\theta}^p+N_m\bar{\theta}^N)^2.}
Computing the fermionic integrals and using the identities \consm ,
\poptp\ and \stuexp\ leads to the following expressions
for the ${\cal I}(i,j;k,l)$~:
\eqn\calIototr{
\eqalign{
{\cal I}(1,2;3,4)=&16N^4(p_1^*)^2p_3^2\cr
{\cal I}(1,3;2,4)=&16\bigl[6{p^4\over N^4}(N_1N_2N_3N_4)^2
+(p_1^*)^2p_3^2N_2^2N_3^2+p_1^2(p_3^*)^2N_1^2N_4^2\cr
&-4{p^2\over N^2}N_1N_2N_3N_4(p_1^*p_3N_2N_3+p_1p_3^*N_1N_4)\bigr]\cr
{\cal I}(1,4;2,3)=&16\bigl[6{p^4\over N^4}(N_1N_2N_3N_4)^2
+(p_1^*)^2p_3^2N_2^2N_4^2+p_1^2(p_3^*)^2N_1^2N_3^2\cr
&+4{p^2\over N^2}N_1N_2N_3N_4(p_1^*p_3N_2N_4+p_1p_3^*N_1N_3)\bigr].\cr}}
The other three ${\cal I}(i,j;k,l)$ can be obtained by taking complex
conjugates of the above results.

Substituting in the ten vectors for the momenta and polarizations 
into the Lorentz invariant expression \otot\ one can show that the
expression \otot\ is equal \otota\calIototr . It is crucial, to obtain this
equality, to include in the polarization vectors the non-zero $\xi^-$ 
part (see equation \xipxim ).
\vskip 17.5pt
Turning to the case with two polarizations in the $X^1$,$X^2$ plane
and two in the $X^3$,$X^4$ plane we have 
\eqn\otthfa{
\int\,d^8\theta\,
\Psi^i_{(12)}\Psi^j_{(12)}\Psi^k_{(34)}\Psi^l_{(34)}
={\xi_i\xi_j^*\tilde{\xi}_k\tilde{\xi}_l^*\over N_iN_jN_l^2}
{\cal I}(i;j|k;l)\quad+\,\,3\,\,{\rm other\,\,permutations},}
where 
\eqn\calIottf{
{\cal I}(i;j|k;l)=-\epsilon\,\,\int\,d^8\theta\,
\prod_{m=j,l}\prod_{n=i,l}(p_m^*\theta^p+N_m\theta^N)^2
(p_n\bar{\theta}^p+N_n\bar{\theta}^N)^2,}
where again $\epsilon$ is a sign factor which is equal to $+1$ if 
$i,j=1,2$ or $3,4$ and $-1$ otherwise. It comes from the 
sign of the $p^+$ momenta in the definitions of the wavefunctions 
\ottffsse .
Evaluating the fermionic integrals and using the identities
\consm , \poptp\ and \stuexp\ one finds that 
\eqn\caliIids{
{\cal I}(i;j|k;l)={\cal I}(i;j|l;k)
\quad{\rm and}\quad
{\cal I}(i;j|k;l)={\cal I}^*(j;i|k;l),}
with the 6 $(=4!/(2!)^2)$ different ways of distributing the two types
of polarization amongst the four wavefunctions being given by
\eqn\calIottfr{\eqalign{
{\cal I}(1;2|3;4)=&16\bigl[-{p^4\over N^4}N_1N_2(N_3^2-4N_3N_4+N_4^2)
+2{p^2\over N^2}(N_3-N_4)(N_1p_1p_3^*-N_2p_1^*p_3)\cr
&\hskip100pt
-{N_1\over N_2}p_1^2(p_3^*)^2-{N_2\over N_1}(p_1^*)^2p_3^2\bigr]\cr
{\cal I}(1;3|2;4)=&16\bigl[{p^4\over N^2}N_1N_3+2p^2p_1^*p_3
+{N^2\over N_1N_3}(p_1^*)^2p_3^2\bigr]\cr
{\cal I}(1;4|2;3)=&16\bigl[{p^4\over N^2}N_1N_4-2p^2p_1^*p_3
+{N^2\over N_1N_4}(p_1^*)^2p_3^2\bigr]\cr
{\cal I}(2;3|1;4)=&16\bigl[{p^4\over N^2}N_2N_3-2p^2p_1^*p_3
+{N^2\over N_2N_3}(p_1^*)^2p_3^2\bigr]\cr
{\cal I}(2;4|1;3)=&16\bigl[{p^4\over N^2}N_2N_4+2p^2p_1^*p_3
+{N^2\over N_2N_4}(p_1^*)^2p_3^2\bigr]\cr
{\cal I}(3;4|1;2)=&16\bigl[-{p^4\over N^4}N_3N_4(N_1^2-4N_1N_2+N_2^2)
-2{p^2\over N^2}(N_1-N_2)(N_4p_1p_3^*-N_3p_1^*p_3)\cr
&\hskip100pt
-{N_4\over N_3}p_1^2(p_3^*)^2-{N_3\over N_4}(p_1^*)^2p_3^2\bigr].\cr}
}
By substituting in the ten vectors for the momenta and polarizations 
into the Lorentz invariant expressions \ottf\ one can show that the 
expressions of \ottf\ are equal to those of \otthfa\calIottfr .
Again it is crucial to include in the polarization vectors the 
non-zero $\xi^-$ part (see equation \xipxim ) to obtain agreement.
\vskip17.5pt
Finally we turn to the case with two polarizations in the $X^1$,$X^2$
plane and two in the $X^5$,$X^6$,$X^7$,$X^8$ subspace. For this case
the fermionic integrals read
\eqn\otfssea{
\int\,d^8\theta\Psi^i_{(12)}\Psi^j_{(12)}\Psi^k_{(5678)}\Psi^l_{(5678)}
={\xi_i\xi_j^*(\xi_k.\xi_l)\over N_1N_2N_3N_4}
{\cal I}(i;j)\quad+\quad{\rm c.c.},}
where 
\eqn\calIij{\eqalign{
{\cal I}(i;j)\delta^{m_km_l}=&-\int\,d^8\theta\,\,(p_j^*\theta^p+N_j\theta^N)^2
(p_i\bar{\theta}^p+N_i\bar{\theta}^N)^2\cr
&\prod_{n=k,l}(p_n^*\theta^p+N_n\theta^N)(\sigma^{m_n})
(p_n\bar{\theta}^p+N_n\bar{\theta}^N).\cr}}
Evaluating the integrals the results are found to be identical to
those of \calIottfr , i.e. we have 
\eqn\calII{
{\cal I}(i;j)={1\over 2}{\cal I}(i;j|k;l),}
where the ${\cal I}(i;j|k;l)$ (which have $k,l\neq i,j$) are given by
the expressions of \calIottfr . 

\listrefs

\bye